\documentclass[10pt,twocolumn, final]{IEEEtran}
\usepackage[dvips,final]{graphicx}
\usepackage{amsmath}
\usepackage{amsthm}
\usepackage{lscape}
\usepackage{latexsym}
\usepackage{amssymb}
\usepackage{bm}
\usepackage{bbm}
\usepackage{algorithm}
\usepackage{todonotes}
\usepackage{color}
\usepackage{multirow}
\usepackage{booktabs}

\usepackage{enumitem}
\usepackage[caption=false,font=footnotesize]{subfig}

\usepackage{algorithm}
\usepackage{algpseudocode}
\usepackage{cite}

\makeatletter
\def\BState{\State\hskip-\ALG@thistlm}
\makeatother

\algnewcommand\algorithmicforeach{\textbf{for each}}
\usepackage{xparse}
\makeatletter

\algdef{SE}[SUBALG]{Indent}{EndIndent}{}{\algorithmicend\ }%
\algtext*{Indent}
\algtext*{EndIndent}

\NewDocumentCommand{\LeftComment}{s m}{%
  \Statex \IfBooleanF{#1}{\hspace*{\ALG@thistlm}}\(\triangleright\) #2}
\makeatother
\algdef{S}[FOR]{ForEach}[1]{\algorithmicforeach\ #1\ \algorithmicdo}

\theoremstyle{definition}

\usepackage{hyperref}

\begin{document}

\title{Deep Video Precoding}
\author{Eirina Bourtsoulatze, Aaron Chadha, Ilya Fadeev, Vasileios Giotsas, and  Yiannis Andreopoulos
\thanks{All authors are with iSIZE Ltd., 3 Falconet Court, 123 Wapping High Street, London, E1W 3NX, United Kingdom,  email: yiannis@isize.co. This paper has been presented in part at the International Broadcasting Conference, IBC 2019, Amsterdam, The Netherlands.}}

\maketitle
\begin{abstract}
Several  groups worldwide are currently investigating how deep learning may advance the state-of-the-art in image and video coding. An open question is how to make deep neural networks work in conjunction with existing (and upcoming) video codecs, such as MPEG H.264/AVC, H.265/HEVC, VVC, Google VP9 and AOMedia AV1, AV2, as well as existing container and transport formats, without imposing any changes at the client side. Such compatibility is a crucial aspect when it comes to practical deployment, especially when considering the fact that the video content industry and hardware manufacturers are expected to remain committed to supporting these standards for the foreseeable future. 

We propose to use deep neural networks as precoders for current and future video codecs and adaptive video streaming systems. In our current design, the core precoding component comprises a cascaded structure of downscaling neural networks that operates during video encoding, prior to transmission. This is coupled with a precoding mode selection algorithm for each independently-decodable stream segment, which adjusts the downscaling factor according to scene characteristics,  the utilized encoder, and the desired bitrate and encoding configuration. Our framework is compatible with all current and future codec and transport standards, as our deep precoding network structure is trained in conjunction with linear upscaling filters (e.g.,  the bilinear filter), which are supported by all web video players. Extensive evaluation on FHD (1080p) and UHD (2160p)\ content and with widely-used H.264/AVC, H.265/HEVC  and VP9 encoders, {as well as a preliminary evaluation with the current test model of VVC (v.6.2rc1)}, shows that coupling such standards with the proposed deep video precoding allows for { 8\% to 52\%} rate reduction under encoding configurations and bitrates suitable for video-on-demand adaptive streaming systems. The use of precoding can also lead to  encoding complexity reduction, which is essential for cost-effective cloud deployment of {complex encoders like H.265/HEVC, VP9 and VVC}, especially when considering the prominence of high-resolution adaptive video streaming. 

\end{abstract}



\section{Introduction}
\label{sec:introduction}

\IEEEPARstart{I} {n just} a few years, technology has completely overhauled the way we consume television, feature films and other prime content. For example, Ofcom reported in July 2018 that there are now more UK subscriptions to Netflix, Amazon and NOW TV than to ‘traditional’ pay TV services.\footnote{https://www.ofcom.org.uk/about-ofcom/latest/media/media-releases/2018/streaming-overtakes-pay-tv} The proliferation of over-the-top (OTT) streaming content has been matched by an appetite for high-resolution content. For example, 50\% of the US homes will have UHD/4K TVs by 2020. At the same time, costs of 4K camera equipment have been falling rapidly. Looking ahead, 8K TVs were introduced at the 2018 CES by several major manufacturers and several broadcasters announced they will begin 8K broadcasts in time for the 2020 Olympic games in Japan. Alas, for most countries, even the delivery of FHD (1080p) content is still plagued by broadband infrastructure problems.

To get round this problem, OTT content providers resort to adaptive streaming technologies, such as Dynamic Adaptive Streaming over HTTP (DASH) and HTTP Live Streaming (HLS), where the streaming server is offering bitrate/resolution ladders via the so-called ``manifest'' file \cite{sodagar2011mpeg,DASH/HLSladder}. This allows the client device to switch to a range of lower resolutions and bitrates when the connection bandwidth does not suffice to support the high-quality/full-resolution video bitstream \cite{sodagar2011mpeg}. In order to produce the bitrate/resolution ladders, high-resolution frames are downscaled to lower resolutions, with the frame dimensions being reduced with decreased bitrate. Within all current adaptive streaming systems, this is done using standard downscaling filters, such as the bicubic filter, and the chosen resolution per bitrate stays constant throughout the content's duration and is indicated in the manifest file. At the client side, if   the chosen bitrate corresponds to low-resolution frames, these frames are upscaled after decoding to match the resolution capabilities of the client's device. Unfortunately, the impact on visual quality from the widely-used bicubic downscaling and the lack of dynamic resolution adaptation per bitrate can be quite severe \cite{katsavounidis2018video}. In principle, this could be remedied by post-decoding learnable video upscaling solutions similar to learnable super-resolution techniques for still images \cite{haris2018deep,LimCVPRW2017EDSR}. However, their deployment requires substantial changes to the client device, which are usually too cumbersome and complex to make in practice (a.k.a., the hidden technical debt of machine learning \cite{sculley2015hidden}). For example, convolutional neural network (CNN) based upscalers with tens of millions of parameters cannot be supported by mainstream CPU-based web browsers that support DASH and HLS video playback.

\begin{figure*}
\centering
\includegraphics[width=1.0\textwidth]{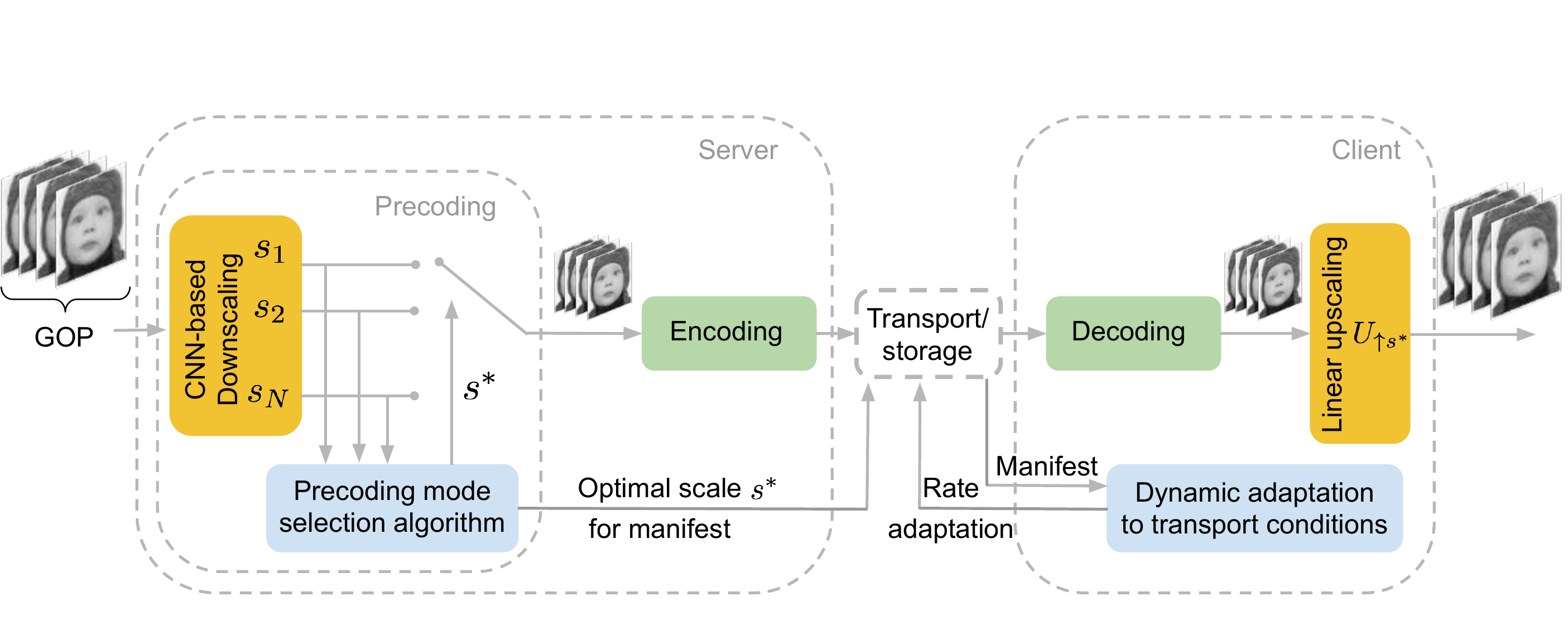}
\caption{Proposed deep video precoding framework. The precoding module performs dynamic resolution adaptation during encoding, prior to streaming. The optimal scale factor is chosen by our adaptive precoding mode selection algorithm which operates per GOP, bitrate and codec and is passed to the server-based manifest file that points all adaptive streaming clients to the available content chunks (GOPs) and their bitrate and resolution settings.}
\label{fig:video_opt_framework}
\end{figure*}

\subsection{Precoding for Video Communications}
\label{sec:intro_precoding}

Precoding has been initially proposed for MIMO wireless communications as the means to preprocess the transmitted signal and perform transmit diversity \cite{WieselTSP2006}. {Precoding is similar to channel equalization, but the key difference is that it shapes (precodes) the signal according to the operation of the utilized receiver \textit{prior} to channel coding. While channel equalization aims to minimize channel errors, a precoder aims to minimize the error in the receiver output.} 

In this paper, we introduce the concept of precoding for adaptive video streaming. As illustrated in Fig. \ref{fig:video_opt_framework}, precoding for video is done by preprocessing the input video frames of each independently-decodable group of pictures (GOP) prior to standard encoding, while allowing a standard video decoder and player to decode and display them without requiring any modifications. {The key idea of precoding is to minimize the distortion at the output of the client's player. To achieve this, we leverage on the support for multiple resolutions at the player side and introduce a multi-scale precoding convolutional neural network (CNN) that progressively downscales input high-resolution frames over multiple scale factors. A mode selection algorithm then selects  the best precoding mode (i.e., resolution) to use per GOP based on the GOP frame content, bitrate and codec characteristics.  The precoding CNN is designed to compact information in such a way that the aliasing and blurring artifacts generated during linear upscaling are mitigated.} This is because, in the vast majority of devices, video upscaling is performed by means of linear filters. Thus, our proposed deep video precoding solution focuses on matching standard video players' built-in linear upscaling filter implementations, such as the bilinear filter\footnote{Despite an abundance of possible upscaling filters, in order to remain efficient over a multitude of client devices, most web browsers only support the bilinear filter for image and video upscaling in YUV colorspace, e.g., see the Chromium source code that uses libyuv (https://chromium.googlesource.com).}.

Our experiments with standard FHD (1080p) and UHD (2160p) test content from the XIPH repository and well-established implementations of H.264/AVC, H.265/HEVC and VP9 encoders show that, by using deep precoding modes for downscaling, we can significantly reduce the distortion of video playback compared to  conventional downscaling filters and fixed downscaling modes used in standard DASH/HLS streaming systems. We further demonstrate through extensive experimentation that the proposed adaptive precoding mode selection achieves {8\% to 52\%} bitrate reduction for FHD and UHD content encoding at typical bitrate ranges used in commercial deployments. An important effect of the proposed precoding is that video encoding can be accelerated, since many GOPs tend to be shrunk by the precoder to only 6\%-64\% of their original size, depending on the selected downscaling factor. This is beneficial when considering cloud deployments of such encoders, especially, in view of upcoming standards with increased encoding complexity.

In summary, our contributions are as follows:
\begin{itemize}
\item the concept of deep precoding for video delivery  is introduced as the means of enhancement of the rate-distortion characteristics of any video codec without requiring any changes on the client/decoder side; 
\item a multi-scale precoding CNN is proposed, which downscales high-resolution frames over multiple scale factors and is trained to mitigate the aliasing and blurring artifacts generated by standard linear upscaling filters;
\item an adaptive precoding mode selection algorithm is proposed, which adaptively selects the optimal resolution prior to encoding.
\end{itemize}

It is important to emphasize that {deep video precoding is a \textit{source encoding} optimization framework carried out at the server side, prior to transmission, in order to optimally ``shape'' the input signal by deriving the best downscaled representation according to input GOP segment, bitrate, codec and upscaling capabilities at the receiver, without considering transport conditions}. Fig. \ref{fig:video_opt_framework} illustrates how our proposed deep video precoding can be used in conjunction with client-driven adaptive streaming systems like the widely-deployed DASH/HLS standards. As shown in Fig. \ref{fig:video_opt_framework}, precoding for each content, codec and target bitrate is independent of how DASH/HLS players will switch between encoding ladders to  adapt the video bitrate to the transport bandwidth, latency and buffer conditions. That is, once the DASH/HLS-compliant stream is produced by our approach for each encoding bitrate and the manifest file is created with the corresponding bitrate and resolution information per video segment, adaptive bitrate streaming, stream caching and stream switching mechanisms that cope with bitrate and channel fluctuations can operate as usual. The key difference in our case is that the client receives, decodes and  upscales bespoke representations created by the proposed precoder. Hence, our proposal for deep video precoding   is done in its entirety during content encoding and remains agnostic to the transport conditions experienced during each individual video bitstream delivery.

\subsection{Paper Organization}
\label{sec:intro_paper_organization}

The remainder of this paper is organized as follows. In Sec. \ref{sec:related_work}, we review related work. The proposed multi-scale deep precoding network architecture design, loss function, and implementation and training details are presented in Sec. \ref{sec:precod_networks}. The precoding mode selection algorithm is presented in Sec. \ref{sec:adaptive_precoding}.  Experimental results are given in Sec. \ref{sec:experiments}, and Sec. \ref{sec:conclusions} concludes the paper.

\section{Related work}
\label{sec:related_work}

Content-adaptive encoding has emerged as a popular solution for quality or bitrate gains in standards-based video encoding. Most commercial providers have already demonstrated content-adaptive encoding solutions, typically in the form of bitrate adaptation based on combinations of perceptual metrics, i.e., lowering the encoding bitrate for scenes that are deemed to be simple enough for a standard encoder to process. Such solutions can also be extended to other encoder parameter adaptations, and their essence is in the coupling of a visual quality profile to a pre-cooked encoder-specific tuning recipe. 

\subsection{ Resolution Adaptation in Image and Video Coding}
{It has been known for some time that reducing the input resolution in image or video coding can improve visual quality for lower bitrates as the encoder operates better at the ``knee'' of its rate-distortion curve \cite{weidmann2012rate}}. Starting from non-learnable designs, Tsaig \textit{et al.} \cite{tsaig2005variable} explored the design of optimal decimation and interpolation filters for block image coders like JPEG and showed that low-bitrate image coding between 0.05 bits-per-pixel (bpp) to 0.35 bpp benefits from such designs.  Kopf \textit{et al.} \cite{kopf2013content} proposed a content-adaptive method, wherein filter kernel coefficients are adapted with respect to image content. Oztireli \textit{et al.} \cite{oztireli2015perceptually} proposed an optimization framework to minimize structural similarity between the nearest-neighbor upsampled low-resolution image and the high-resolution image. Recently, Katsavounidis \textit{et al.} \cite{katsavounidis2018video, bampis2018towards} proposed the notion of the dynamic optimizer in video encoding: each scene is downscaled to a range of resolutions  and is subsequently compressed to a range of bitrates. After upscaling to full resolution, the convex hull of bitrates/qualities is produced in order to select the best operating resolution for each video segment. Quality can be measured with a wide range of metrics, ranging from simple peak signal to noise ratio (PSNR) to complex fusion-based metrics like the video multimethod assessment fusion (VMAF) metric \cite{li2016toward}. While Bjontegaard distortion-rate (BD-rate) \cite{bjontegaard2001calculation} gains of 30\% have been shown in experiments for H.264/AVC and VP9, the dynamic optimizer requires very significant computational resources, while it still uses non-learnable downscaling filters.  {A learnable CCN-based downscaling method for image compaction was proposed by Li \textit{et al.} \cite{LiTIP2019}, where the downscaling CNN is trained jointly with either a fixed linear upscaling filter or a trainable upscaling CNN. While that work also explored the concept of learnable downscaling, it is designed for fixed-ratio downscaling and does not provide content, bitrate and codec adaptivity as our proposed deep precoding approach. However, it forms an important learnable downscaling framework that we use as one of the benchmarks for our work. }

\subsection{ Super-resolution Methods}

Overall, while the above methods have shown the possibility of rate saving via image and video downscaling, they have not managed to {significantly} outperform classical bicubic downscaling within the context of practical encoding. This has led most researchers and practitioners to conclude that downscaling with bicubic or Lanczos filters is the best approach, and instead the focus has shifted on upscaling solutions at the client (i.e., decoder) side that learn to recover image detail assuming such downscaling operators. {For example, Georgis et al. \cite{georgis2016reduced} proposed backprojection-based upscaling that is tailored to Gaussian kernel downscaling and showed that such approaches can be beneficial for FHD and UHD/4K encoding with H.264/AVC and HEVC up to 10mbps and 5mbps, respectively. The majority of recent works in this field consider CNN-based upscaling methods. This has been largely motivated by the success of deep CNN architectures for single image super-resolution}, that have set the state-of-the-art, with recent architectures like VDSR \cite{kim2016accurate}, EDSR \cite{lim2017enhanced}, FSRCNN \cite{dong2016accelerating}, DRCN \cite{kim2016deeply} and DBPN \cite{haris2018deep} achieving several dB higher PSNR in the luminance channel of standard image benchmarks for lossless image upscaling. Thus,  Afonso \textit{et al.} propose a spatio-temporal resolution adaptation where a CNN-based super-resolution model is used to reconstruct full-resolution content \cite{afonso2018video}. Li \textit{et al.} \cite{LiTCSVT2018Intra} introduce the block adaptive resolution coding framework for intra frame coding, where each block within a frame is either downscaled or coded at original resolution and then upscaled with a trained CNN at the decoder side. This concept was later extended to include P and B frames as well \cite{LinTCSVT2018HEVC}. Differently from the previous methods that operate in the pixel domain, Liu \textit{et al.} perform down and upsampling in the residue domain and design upsampling CNN for residue super resolution (SR) with the help of the motion compensated prediction signal \cite{LiuVCIP2018}. However, regardless of the domain where they operate, the common principle of all these works is that they use hand-crafted filters for downscaling and perform upscaling at the decoder side using codec-tailored CNN models integrated in the coding standard which requires modifications to the codec.

\subsection{Neural-network Representation and Coding Methods }

While most research efforts have focused on learning optimal upscaling filters \cite{LiuARXIV2019review}, inspired by the success of autoencoders for image compression \cite{TheisICLR2017, RippelICML2017}, some recent works revise the problem of joint downscaling and upscaling using deep CNN-based methods. Shocher \textit{et al.} \cite{shocher2018zero} recently proposed an upscaling method using deep learning, which trains an image-specific CNN with high and low-resolution pairs of patches extracted from the test image itself.  Weber \textit{et al.} \cite{weber2016rapid} used convolutional filters to preserve important visual details, and Hou \textit{et al.} \cite{hou2017deep} recently proposed a deep perceptual loss based method. Kim \textit{et al.} \cite{kim2018task} proposed an image downscaling/upscaling method using a deep convolutional autoencoder, and achieved state-of-the-art results on standard image benchmarks. An end-to-end image compression framework for low bitrates compatible with existing image coding standards  was introduced in \cite{JiangTCSVT2018}. It comprises two CNNs: one for image compaction prior to encoding and one for post-processing after decoding. Adaptive spatio-temporal decomposition prior to encoding, followed by CNN-based spatio-temporal upscaling after decoding was proposed by Afonso \textit{et al.} and was validated with H.265/HEVC encoding \cite{afonso2018video}. Finally, Wave One recently proposed video encoding with deep neural networks \cite{rippelARXIV2019videocompression} and demonstrated quality gains against a conventional video encoder without B frames, and focusing on very-high bitrate encoding (20mbps or higher for FHD). While these are important achievements, most of these proposals are still outperformed by post-2013 video encoders, like HEVC and VP9, when utilized with their most advanced video buffering verifier (VBV) encoding configurations and appropriate constant rate factor tuning \cite{asan2018optimum}. In addition, all these proposals require advanced GPU capabilities on both the client and the server side that cannot be supported by existing video players as they break away from current standards. Therefore, despite the significant advances that may be offered by all these methods in their future incarnations, they do not consider the stringent complexity and standards compatibility constraints imposed when dealing with adaptive video streaming under DASH or HLS-compliant clients like web browsers. Our work fills this gap by offering deep video precoding as the means to optimize existing video encoders with the  entirety of the precoding process taking place on the server side and not requiring any change in the video transport, decoding and display side.

\section{Multi-scale precoding networks}
\label{sec:precod_networks}

While any downscaling method can be used with the video precoding framework of  Fig. \ref{fig:video_opt_framework}, to  enhance the performance of our proposal in a data-driven manner, we introduce a multi-scale precoding neural network. The  precoding network comprises a series of CNN precoding blocks that progressively downscale high resolution (HR) video frames over multiple scale factors. We design the precoding CNN to compact information such that a standard linear upscaler at the video player side will be able to recover in the best possible way. This is the complete opposite of  recent image upscaling  architectures that assume simple bicubic downscaling and an extremely complex super-resolution CNN architecture at the video player side. For example, EDSR \cite{LimCVPRW2017EDSR}  comprises over 40 million parameters and would be highly impractical on the client side for 30-60 frame-per-second (fps) FHD/UHD video.  

In the following subsections, we describe the design of the proposed multi-scale precoding networks, including the network architecture, loss function, and details of implementation and training.

\subsection{Network Architecture}
\label{sec:network_arch}

\begin{figure}
\centering
\includegraphics[width=0.5\textwidth]{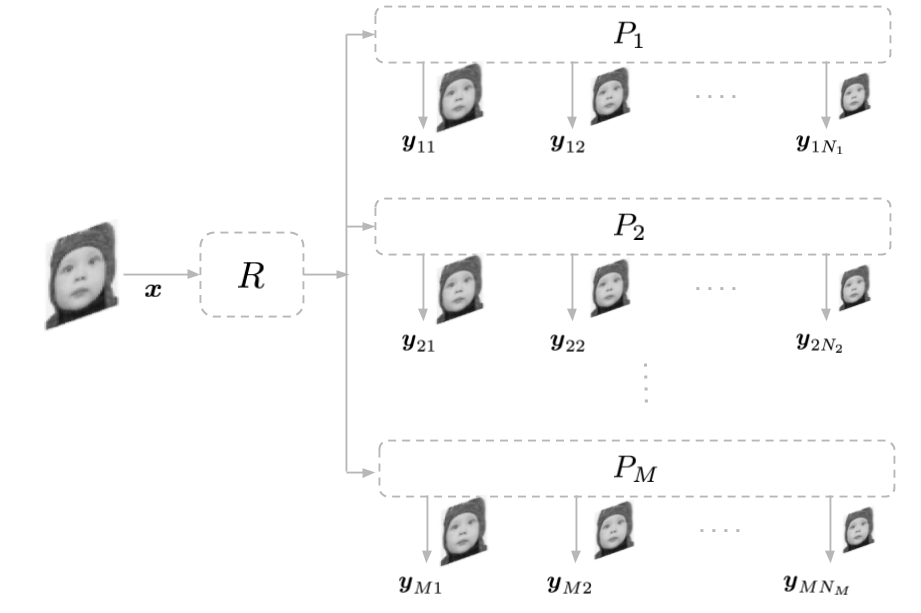}
\caption{The architecture of our multi-scale precoding network for video downscaling, comprising a root mapping $R$ and  precoding streams $P_1, P_2, \dots P_M$. The luminance frame of  each input video frame is downsampled to multiple lower resolutions by the precoding network at the server via the precoding streams. }
\label{fig:precoding_network}
\end{figure}

The overall architecture of the precoding network is depicted in Fig. \ref{fig:precoding_network}. It consists of a ``root'' mapping $R$ followed by $M$ parallelized precoding streams $P_{m}$. The network progressively downscales individual luminance frames $\boldsymbol{x} \in \mathbb{R}^{H \times W}$(where $H$ and $W$ are the height and width, respectively) over the scale factors in $\boldsymbol{S}$.  Considering that the human eye is most sensitive to luma information, we intentionally process only the luminance (Y) channel with the precoding network and not the chrominance (Cb, Cr) channels, in order to avoid unnecessary computation. Dong \textit{et al.} \cite{dong2016image} support this claim empirically and, additionally, find that training a network on all three channels can actually worsen performance due to the network falling into a bad local minimum. We also note that this permits for chroma subsampling (e.g., YUV420) as the chrominance channels (Cb,Cr)  are downscaled independently using the standard bicubic filter.

\subsubsection{Root mapping}
\label{sec:root_mapping}
\begin{figure}
        \begin{center}
                \includegraphics[width=0.5\textwidth]{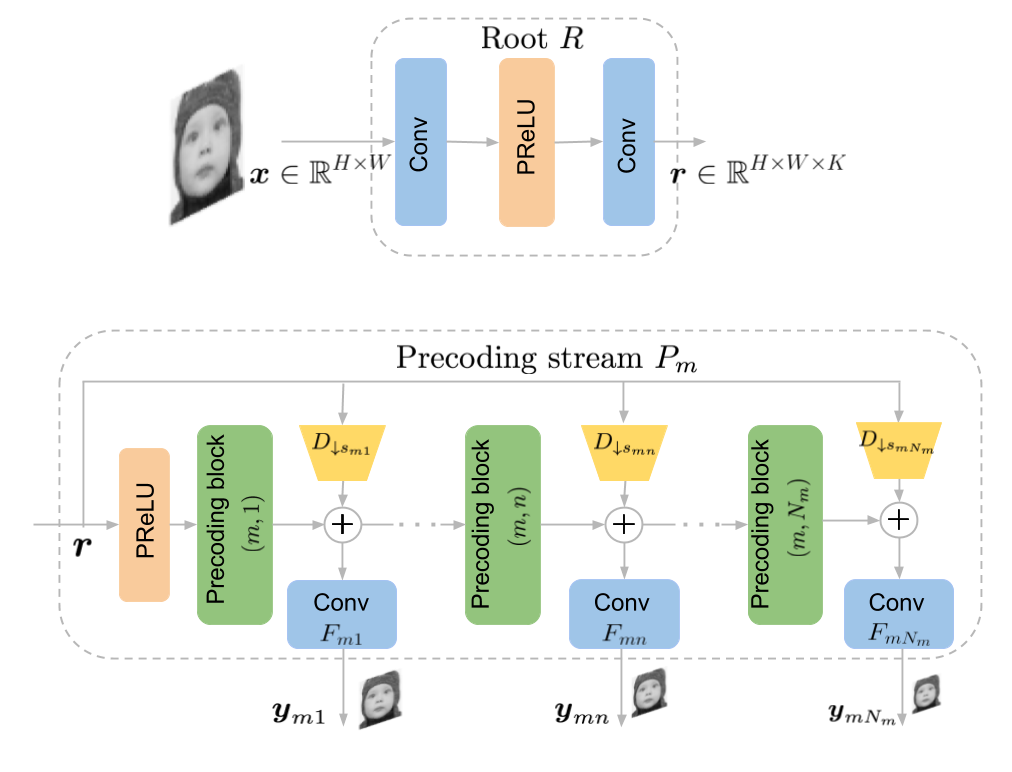}
        \end{center}
                \caption{Root mapping $R$ and $m$-th precoding stream $P_m$. The root mapping extracts high-dimensional feature maps $\boldsymbol{r}$ and is shared by all precoding streams. The precoding stream $P_m$ contains a sequence of precoding blocks and progressively downsamples the input high-resolution frames over a set of $N_m$ scale factors. }
        \label{fig:root_and_precoding_stream}
\end{figure}

 The root mapping $R$, illustrated in Fig \ref{fig:root_and_precoding_stream}, comprises two convolutional layers and extracts a set of high-dimensional feature maps $\boldsymbol{r} \in \mathbb{R}^{H \times W \times K}$ from the input $\boldsymbol{x}$, where $K$ is the number of output channels of the root mapping. The root mapping $R$ constitutes less abstract features (such as edges) that are common amongst all precoding streams and scale factors.  Therefore, this module is shared between all precoding streams, which helps in reducing complexity.

\subsubsection{Precoding stream}
\label{sec:precoding_stream}
The extracted feature maps $\boldsymbol{r}$ are passed to the precoding streams. As depicted in Fig. \ref{fig:root_and_precoding_stream}, a precoding stream $P_{m}$  comprises a sequence of $N_m$ precoding blocks, which progressively downscale the input over a subset of $N_m$ scale factors, $\boldsymbol{S}_m = \{{s_{m1}}, {s_{m2}}, \dots ,{s_{mN_m}}\} \subseteq \boldsymbol{S}$, where $1 < s_{m1}   < s_{m2}  < \dots <s_{mN_m}  $. The  allocation of scale factors to a precoding stream is done in such a way that: \textit{(i)} complexity is shared equally between streams for efficient parallel processing 
and {\textit{(ii)} the ratio of most of the consecutive pairs of scales $s_{m(n-1)}, s_{mn}$ within a stream is constant, i.e., $\alpha_{mn}=s_{mn}/s_{m(n-1)} =\alpha_{m}\in\mathbb{Z}^{+}$.}
Such construction of precoding streams exploits the inter-scale correlations among similar scales and enables parameter sharing within each precoding stream, thus further reducing the computational complexity. Most importantly, it renders our network amenable to standard encoding resolution ladders, such as those used in DASH/HLS streaming \cite{DASH/HLSladder}.

Given a precoding stream $P_m$, the $n$-th constituent precoding block receives a function of the output map  $\boldsymbol{p}_{m(n-1)_{}} \in \mathbb{R}^{H/s_{m(n-1)} \times W/s_{m(n-1)} \times K}$ from the preceding block and outputs embedding $\boldsymbol{p}_{mn} \in \mathbb{R}^{H/{s_{mn}} \times W/{s_{mn}} \times K}$. Notably,  we utilize a global residual learning strategy, where we use a skip connection and perform a pixel-wise summation between the root feature maps $\boldsymbol{r}$ (pre-activation function and after linear downscaling to the correct resolution with a linear downscaling operator $D_{\downarrow{s_{mn}}}$) and   $\boldsymbol{p}_{mn}$.  Similar global residual learning implementations have also been adopted by SR models \cite{lai2017deep,tai2017image,tai2017memnet}. In our case,  our precoding stream effectively follows a pre-activation configuration \cite{he2016identity} without batch normalization.  We find empirically that convergence during training is generally faster with  global residual learning, as the precoding blocks only have to learn the residual map to remove distortion introduced by downscaling operations.

The resulting downscaled feature map is finally mapped by a single convolutional layer $F_{mn}$ to  $\boldsymbol{y}_{mn}$. {Importantly,  the sequence of precoding blocks only operates in a higher dimensional feature space without any heavy bottlenecks. It is then the job of convolutional layers $F_{mn}$ to aggregate block outputs into single channel representations of the input $\boldsymbol{x}$, downscaled by a factor of $s_{mn}$. For $N_m$ precoding blocks and set of  $N_m$ scale factors $\boldsymbol{S}_m$, the embedding stream outputs a set of $N_m$ corresponding downscaled representations $\boldsymbol{Y}_m=\{\boldsymbol{y}_{m1}, \boldsymbol{y}_{m2}, \dots ,\boldsymbol{y}_{mN_m}\}$ of the input $\boldsymbol{x}$.  

The output activations in $\boldsymbol{Y}_m$  are clipped (rescaled) between the minimum and maximum pixel intensities and each representation can, thus, be  passed to the codec  as a   downscaled low resolution (LR) frame. These frames can then be individually upscaled to the original resolution using a  linear  upscaling $U_{\uparrow{s}}$ on the client side, such as bilinear, lanczos or bicubic, where ${\uparrow{s}}$ indicates upscaling by scale factor $s $. We refer to the upscaled frame generated from the downscaled frame $\boldsymbol{y}_{mn}$ as $\hat{\boldsymbol{x}}_{mn}$ and denote the set of $N_m$ upscaled representations of $\boldsymbol{x}$ as $\hat{\boldsymbol{X}}_m = \{ \hat{\boldsymbol{x}}_{m1}, \hat{\boldsymbol{x}}_{m2}, \dots ,\hat{\boldsymbol{x}}_{mN_m}\}$.

 \begin{figure}
\centering
\includegraphics[width=0.5\textwidth]{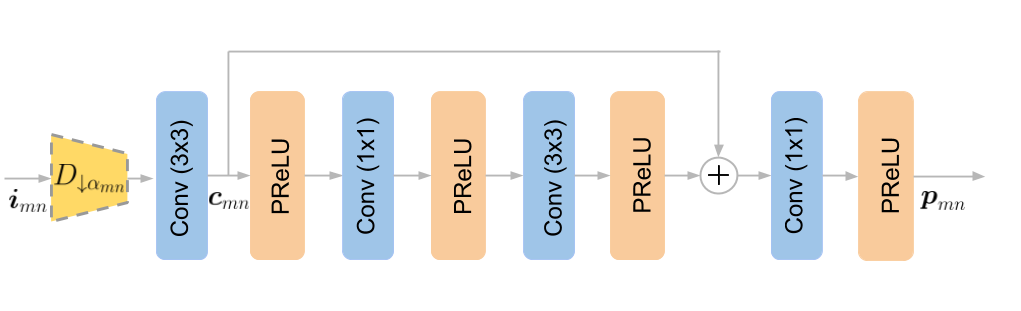}
\caption{Precoding block design, comprising a series of $3 \times 3$ and 1 $\times$ 1 convolutions.  The linear downscaling operation $D_{\downarrow{\alpha_{mn}}}$ is only performed when the downscaling to the target resolution cannot be achieved via stride in the first convolutional layer.  The linear mapping learned from the first layer (pre-activation function) is passed to the output of the second $3 \times 3$ convolutional layer (post-activation function)  with a skip connection and pixel-wise summation.}
\label{fig:precoding_block}
\end{figure}


\subsubsection{Precoding block}
\label{sec:precod_block}

Our precoding block, which constitutes the primary component of our network, is illustrated in Fig. \ref{fig:precoding_block}. The precoding block consists of  alternating $3 \times 3$ and $1 \times 1$ convolutional layers,  where each layer is followed by a parametric ReLU (PReLU) \cite{HeICCV2015PReLU} activation function.  The $1 \times 1$ convolution is used as an efficient means for channel reduction in order to reduce the overall number of multiply-accumulates (MACs) for computation.   

The $n$-th precoding block in the $m$-th precoding stream is effectively responsible for downscaling the original high resolution frame by a factor of ${s_{mn}}$. In order to maintain low complexity, it is important to downscale to the target resolution as early as possible. Therefore, we group all downsampling operations with the first convolutional layer in each precoding block. Denoting the input to the precoding block  as $\boldsymbol{i}_{mn}$, the output of the first convolutional layer   $C_{mn}$ of the precoding block as ${\boldsymbol{c}}_{mn}$ (as labelled in Figure \ref{fig:precoding_block}) and spatial stride as $k$:

\begin{equation}
 \boldsymbol{c}_{mn}= 
\begin{cases}
    C_{mn}(\boldsymbol{i}_{mn};k=\alpha_{mn}),  & \text{if } \alpha_{mn}  \in \mathbb{Z}^+ \\
     C_{mn}(D_{\downarrow \alpha_{mn}}(\boldsymbol{i}_{mn});k=1),            &   \text{otherwise}
\end{cases}
\end{equation}

In other words, downscaling is implemented in the first convolutional layer with a  stride if $\alpha_{mn}$ is an integer; otherwise we use a preceding  linear downscaling operation  $D_{\downarrow \alpha_{mn}}$ (bilinear or bicubic).

The aim of the precoding block is to reduce aliasing artifacts in a  data-driven manner. Considering that the upscaling  operations are  linear and, therefore, heavily constrained, the network is asymmetric and the precoding structure can not simply learn to pseudo invert the upscaling.   If that were to be the case, it would simply result in a traditional linear anti-aliasing filter, i.e., a lowpass filter that removes the high frequency components of the image in a globally-tuned manner, such that the image can be properly resolved.  Removing the high frequencies from the image  leads to blurring.  Adaptive deblurring is an inverse problem, as there is typically limited information in the blurred image to uniquely determine a viable input. As such, within our proposed precoding structure, we can respectively model the anti-aliasing and deblurring processes as a function composition of linear and non-linear mappings.  As illustrated in Fig. \ref{fig:precoding_block}, this is implemented with a skip connection between linear and non-linear paths, as utilized in ResNet \cite{he2016deep} and its variants, again following the pre-activation structure. In order to
ensure that the output of the non-linear path has full range ($-\infty$, $\infty$), we can initialize the
final PReLu before the skip connection such that it approximates an identity function.

\subsection{Loss Function}
\label{sec:loss_func}
Given the luminance channel of a ground-truth frame $\boldsymbol{x} \in \mathbb{R}^{H \times W}$    (with Y ranging between 16-235 as per ITU-R BT.601 conversion), our goal is to learn the parameters for the root $R$ and all precoding streams $P_1, P_2, \dots P_M$.  We denote the root module parameters as $\boldsymbol{\theta}$ and the parameters of the $m$-th precoding stream as $\boldsymbol{\phi}_m$.  For the $m$-th precoding stream with downscaling over $N_m$ scale factors and $I$ training samples per batch, the composite loss function $\mathcal{L}_m$ can be defined as: 

\begin{equation}
\begin{aligned}
\mathcal{L}_m(\hat{\boldsymbol{X}}_m,\boldsymbol{x};\boldsymbol{\theta},\boldsymbol{\phi}_m) = & \frac{1}{I}\sum^{I}_{i=1}\sum^{N_m}_{n=1} \Big{(}\left\vert\hat{\boldsymbol{x}}_{mn}^{(i)}-\boldsymbol{x}^{(i)}   \right\vert  \\ & + \lambda \left\vert \nabla \hat{\boldsymbol{x}}_{mn}^{(i)}-\nabla \boldsymbol{x}^{(i)} \right\vert\Big{)}
 \end{aligned}
 \label{eq:loss_func_m}
\end{equation}

The first term represents the L1 loss between each generated upscaled frame $\hat{\boldsymbol{x}} _{mn}=U_{\uparrow s_{mn}}(\boldsymbol{y}_{mn})$ and the ground-truth frame $\boldsymbol{x}$, summed over all $N_m$ scales, where $\boldsymbol{y}_{mn} =  F_{m,n} \circ P_{m,\downarrow s_{mn}} \circ R  (\boldsymbol{x};\boldsymbol{\theta},\boldsymbol{\phi}_m)$ and $P_{m,\downarrow s_{mn}}$ is the part of the $m$-th precoding stream that includes all precoding blocks up to the $n$-th block and is responsible for downscaling the input feature maps $\boldsymbol{r}$ by the scale factor $s_{mn}$.  The second term represents the L1 loss between the first order derivatives of the generated high resolution and ground-truth frames. As the first order derivatives correspond to edge extraction, the second term acts as an edge preservation regularization for improving perceptual quality.  {We set the weight coefficient $\lambda$ to 0.5 for all experiments; empirically, this was found to produce the best visual quality in output frames.} {Contrary to recent work \cite{kim2018task,LiTIP2019}, we do not add a loss function constraint between the downscaling and upscaling as our upscaling is only linear and, therefore, already heavily constrains the downscaled frames. Most importantly, we do not include the codec in the training process and train end-to-end (i.e., without encoding/transcoding), such that the model does not learn codec specific dependencies and is able to generalize to multiple codecs.} Finally, as we train the parallelized multi-scale precoding network over all streams synchronously,  our final loss function is the summation of \eqref{eq:loss_func_m} over all $M$ precoding streams:  

\begin{equation}
\begin{aligned}
\mathcal{L}(\hat{\boldsymbol{X}},\boldsymbol{x};\boldsymbol{\theta},\boldsymbol{\phi}) =\sum_{m=1}^M \mathcal{L}_m(\hat{\boldsymbol{X}}_m,\boldsymbol{x};\boldsymbol{\theta},\boldsymbol{\phi}_m)
\end{aligned}
\label{eq:loss_func}
\end{equation}

\noindent where $\hat{\boldsymbol{X}} = \hat{\boldsymbol{X}}_1 \cup\ \hat{\boldsymbol{X}}_2 \cup \dots \cup \hat{\boldsymbol{X}}_M$ and $\boldsymbol{\phi} = \{\boldsymbol{\phi}_1, \boldsymbol{\phi}_2,\dots,\boldsymbol{\phi}_M\}$.

\subsection{Implementation and Training Details}
\label{sec:impl_training}

In our proposed multi-scale precoding network, we initialize all kernels using the method of Xavier intialization \cite{glorot2010understanding}. We use PReLU as the activation function, as indicated in Fig. \ref{fig:root_and_precoding_stream} and \ref{fig:precoding_block}.  We use zero padding to ensure that all layers are the same size and downscaling is only controlled by a downsampling operation such as a stride or a linear downscaling filter. The root mapping $R$  comprises a single $3 \times 3$ and $1 \times 1$ convolutional layers. We set the number of channels  in all $1 \times 1$ and  $3 \times 3$ convolutional layers to 4 and  8, respectively {(excluding $F_{m,n}$, which uses a kernel size of $3 \times 3$ but with only a single output channel)}. 

Our final parallelized implementation comprises three precoding streams $P_1, P_2$ and $P_3$, with the set of  scale factors  $\boldsymbol{S}\setminus\{1\}$ partitioned into three subsets: $\boldsymbol{S}_1 =\{4/3, 2, 4\}$, $\boldsymbol{S}_2 =\{3/2, 3, 6\}$ and  $\boldsymbol{S}_3 =\{5/4, 5/2\}$. Collectively, these include all representative scale factors used in DASH/HLS streaming systems \cite{DASH/HLSladder}, as well as additional scale factors that offer higher flexibility in our adaptive mode selection algorithm.  In terms of complexity, for a $1920 \times 1080 \times 4$ -dimensional feature map (assuming no downscaling), a single precoding block requires approximately only 1.33G MACs for downscaling and 640 parameters. Our final implementation requires only 3.38G MACs and 5.5K parameters over all scales per FHD input frame  ($1920 \times 
 1080$), including root mapping and all precoding streams.

We train the root module and all precoding streams  end-to-end with  linear upscaling, without the codec, on images from the DIV2K \cite{agustsson2017ntire} training set. We train all models with the Adam optimizer with a batch size of 32 for 200k iterations.  The initial learning rate is set as 0.001 and decayed by a factor of 0.1 at 100k iterations. We use data augmentation during training, by randomly flipping the images and train with a $120 \times 120$     random crops extracted from the DIV2K images. All experiments were conducted in Tensorflow on NVIDIA K80 GPUs.  We do not use Tensorflow's built-in linear image resizing functions and rewrite all linear upscaling/downscaling functions from scratch, such that they match standard FFmpeg and OpenCV implementations.

\section{Adaptive video precoding}
\label{sec:adaptive_precoding}


Given the trained multi-scale precoding network, the  mode selection algorithm operates on a per GOP, bitrate and codec basis. The goal of mode selection is to determine the optimal precoding scale factor for each GOP.  While one can use  operational rate-distortion (RD)\ models for H.264/AVC or H.265/HEVC \cite{MaTCSVT2005} for evaluation of the best precoding modes, such models cannot encapsulate the complex and sequence-dependent RD\ behavior of each codec preset. On the other hand, exhaustive search solutions  like the Netflix dynamic optimizer \cite{katsavounidis2018video} require numerous highly-complex encodings per bitrate and scale factor. This makes them impractical for high-volume/low-cost encoding systems. Our approach provides a middle ground between these two opposites by computing   operational rate-distortion characteristics of the video encoder for each precoding mode and video segment in an efficient manner.

The precoding mode selection algorithm is outlined in Algorithm \ref{alg:algo} and  comprises three steps. The first step is to obtain the rate-distortion characteristics of each precoding mode (scale). Let $\boldsymbol{S} = \{s_1, s_2, \dots, s_N\}$ be the complete  set of scale factors that may include the native (i.e., full) resolution. Each GOP segment $\boldsymbol{g}$ is precoded into all possible scales in $\boldsymbol{S}$ with the multi-scale precoding network. Let $\boldsymbol{h}_i$ denote the precoded version of GOP $\boldsymbol{g}$  using scale factor $s_i \in \boldsymbol{S}$. All  precoded versions of  GOP $\boldsymbol{g}$  are then encoded with the video encoder's preset and encoding parameters. As such, per scale factor $s_i$, we obtain \textit{a single} GOP encoding, which, after decoding and upscaling to the native resolution, produces \textit{a single} rate-distortion point $(R_i,D_i)$ on the RD plane.\footnote{We use mean squared error (MSE) to measure distortion. Even though other distortion measures could be used, MSE is fast to compute and is automatically provided by all encoders.} This provides significant reduction of the required encodings versus approaches like the Netflix dynamic optimizer, which needs a convex hull of \textit{several} rate-distortion points per scale factor $s_i$, i.e., several encodings per scale factor \cite{katsavounidis2018video}. To accelerate this step even further and avoid unnecessary computational overhead, we introduce a ``footprinting'' process: instead of encoding all frames of the GOP, we perform selective encoding of only a few frames, e.g., keeping only every $n$-th frame in the GOP. This significantly speeds up the initial encoding step, especially if multiple precoding modes are considered, as $n$ times fewer frames need to be precoded and encoded per  scale factor.
We store  the set of  tuples $\boldsymbol{R}=\{(R_i,D_i,\boldsymbol{h}_i,s_i)\}^{\left\vert \boldsymbol{S} \right\vert}_{i=1}$.

Once all the RD points are obtained, we start the pruning process. First, we eliminate all precoding modes, whose RD points do not provide for a monotonically decreasing RD curve. That is, for every precoding mode $i$, if there exists a precoding mode $j$, such that $R_j \leq R_i$ and $D_j < D_i$, the $s_i$ precoding mode is pruned out. If, after this elimination procedure, the number of remaining precoding modes is greater than two, we further prune out precoding modes by eliminating those modes whose RD points do not lie on the convex hull of the remaining RD points \cite{OrtegaSIGPROMAG1998},\cite{katsavounidis2018video}. We refer to the set of pruned tuples as $\boldsymbol{R}_\mathrm{pruned} \subseteq \boldsymbol{R}$. After the pruning stage, we re-encode the remaining precoded GOP representations $\boldsymbol{h}_i$ using constant bitrate (CBR) encoding. The bitrate used for the CBR encoding is equal to the average of the bitrates of all RD points remaining after the elimination step. After decoding and upscaling to the native resolution, we obtain a new set of tuples, $\boldsymbol{R}'_\mathrm{pruned}$, corresponding to the CBR encoding. This final step of the algorithm essentially remaps the RD points that remain after the pruning process to a common bitrate value, which is enforced by the CBR encoding. The optimal precoding mode $s^*$ is then selected as the mode providing the lowest distortion among this new set of remapped RD points. 
  
 \begin{algorithm}
\caption{: Algorithm for adaptive mode selection}
 \textbf{Input}: 
 GOP segment  $\boldsymbol{g}$,  complete set of scale factors $\boldsymbol{S}$ \\
 \textbf{Output}:
  optimal precoding mode $s^*$, optimal precoded version  $ \boldsymbol{h^*}$  of   GOP segment  $\boldsymbol{g}$
\begin{algorithmic}[1]
\Procedure{ModeSelection}{$\boldsymbol{g},\boldsymbol{S}$}
\Indent
\LeftComment{Step 1: Extract RD points for all precoding modes}
\EndIndent
\State $\boldsymbol{R} \leftarrow \{\}$
\ForEach{$s \in \boldsymbol{S}$}      
    \State  $\boldsymbol{h} \leftarrow$ \Call{Precode}{$\boldsymbol{g}, s$} 
    \State  {$(\boldsymbol{e}, R)$} $\leftarrow$ \Call{Encode}{$\boldsymbol{h}$, preset, params} 
    \State  {$\hat{\boldsymbol{h}}$} $\leftarrow$ \Call{Decode}{$\boldsymbol{e}$} 
    \State {$\hat{\boldsymbol{g}}$} $\leftarrow$ \Call{Upscale}{$\hat{\boldsymbol{h}}$}     
    \State {$D \leftarrow \left\Vert \hat{\boldsymbol{g}} -  \boldsymbol{g} \right\Vert^2$}
    \State {$\boldsymbol{R} \leftarrow \boldsymbol{R} \cup \{(R,D,\boldsymbol{h},s)\}$}
\EndFor
\Indent
\LeftComment{Step 2: Prune out RD points}
\EndIndent
\State {$\boldsymbol{R}_\mathrm{sorted} \leftarrow((R_i,D_i,\boldsymbol{h}_i,s_i) \in \boldsymbol{R} \vert R_i>R_{i-1})_{i=1}^{\left\vert \boldsymbol{S} \right\vert}$ }
\State {$\boldsymbol{R}_\mathrm{pruned} \leftarrow \{\boldsymbol{R}_\mathrm{sorted}[1]\}$; $D_\mathrm{ref} \leftarrow D_1$}
\If  {$\left\vert \boldsymbol{S}\right\vert >1$}
        \For {$i=2$ \textbf{to} $\left\vert \boldsymbol{S}\right\vert$}
        \State {$(R_i,D_i) \leftarrow\boldsymbol{R}_{\mathrm{sorted}}[i]$}

\If {$ D_i < D_\mathrm{ref} $}
\State $\boldsymbol{R}_\mathrm{pruned} \leftarrow \boldsymbol{R}_\mathrm{pruned} \cup \{(R_i, D_i, \boldsymbol{h}_i, s_i)\}$
\State $D_{\mathrm{ref}} \leftarrow D_i$
\EndIf

\EndFor
\EndIf
\If {$\left\vert  \boldsymbol{R}_\mathrm{pruned}  \right\vert >2$ }
        \State $\boldsymbol{R}_\mathrm{pruned} \leftarrow \Call{ConvexHull}{\boldsymbol{R}_\mathrm{pruned}} $
\EndIf
\Indent
\LeftComment{Step 3: Re-encode remaining RD points with CBR }
\EndIndent
\State $\boldsymbol{R}'_\mathrm{pruned} \leftarrow \{\}$
\ForEach {$(R,D,\boldsymbol{h},s) \in \boldsymbol{R}_\mathrm{pruned}$}

\State  {$(\boldsymbol{e}, R)$} $\leftarrow$ \Call{Encode}{$\boldsymbol{h}$, preset, params, `CBR'} 
\State  {$\hat{\boldsymbol{h}}$} $\leftarrow$ \Call{Decode}{$\boldsymbol{e}$} 
\State {$\hat{\boldsymbol{g}}$} $\leftarrow$ \Call{Upscale}{$\hat{\boldsymbol{h}}$}     
\State {$D \leftarrow \left\Vert \hat{\boldsymbol{g}} -  \boldsymbol{g} \right\Vert^2$}
\State {$\boldsymbol{R}'_\mathrm{pruned} \leftarrow \boldsymbol{R}'_\mathrm{pruned} \cup \{(R,D,\boldsymbol{h},s)\}$}
\EndFor
\State {$(R^*,D^*,\boldsymbol{h}^*, s^*) \leftarrow \min _{D}(\boldsymbol{R}'_\mathrm{pruned})$}

\EndProcedure

\end{algorithmic}
\label{alg:algo}
\end{algorithm}

\begin{figure*}
\centering

        \subfloat[]{\includegraphics[width=0.24\textwidth]{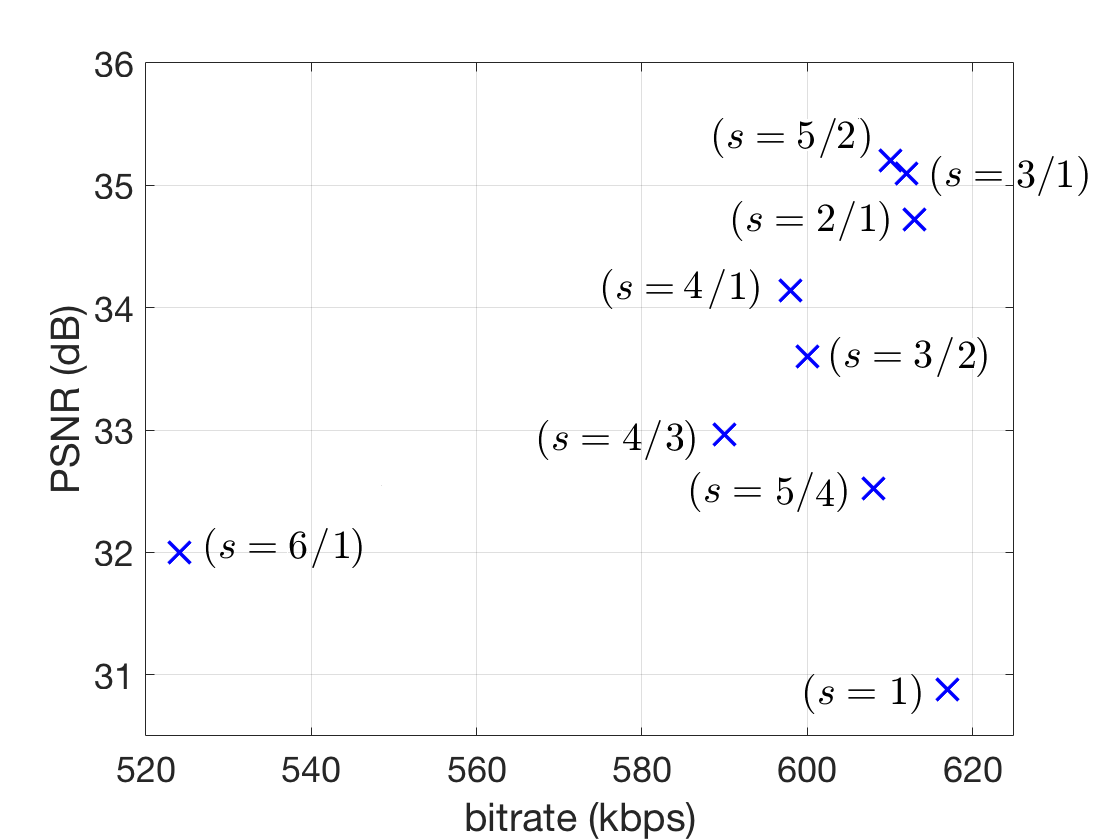}\label{fig:footprinting}} ~ 
        \subfloat[]{\includegraphics[width= 0.24\textwidth]{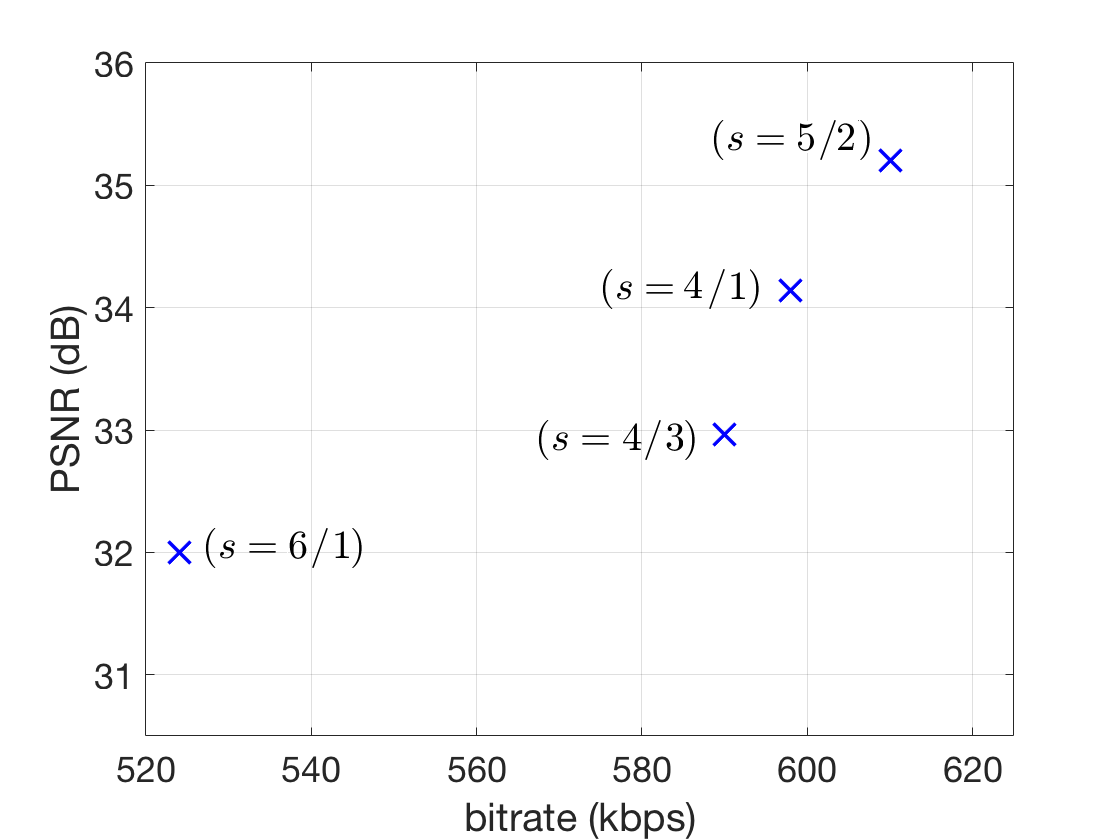}\label{fig:elimination_monotonicity}} ~
        \subfloat[]{\includegraphics[width= 0.24\textwidth]{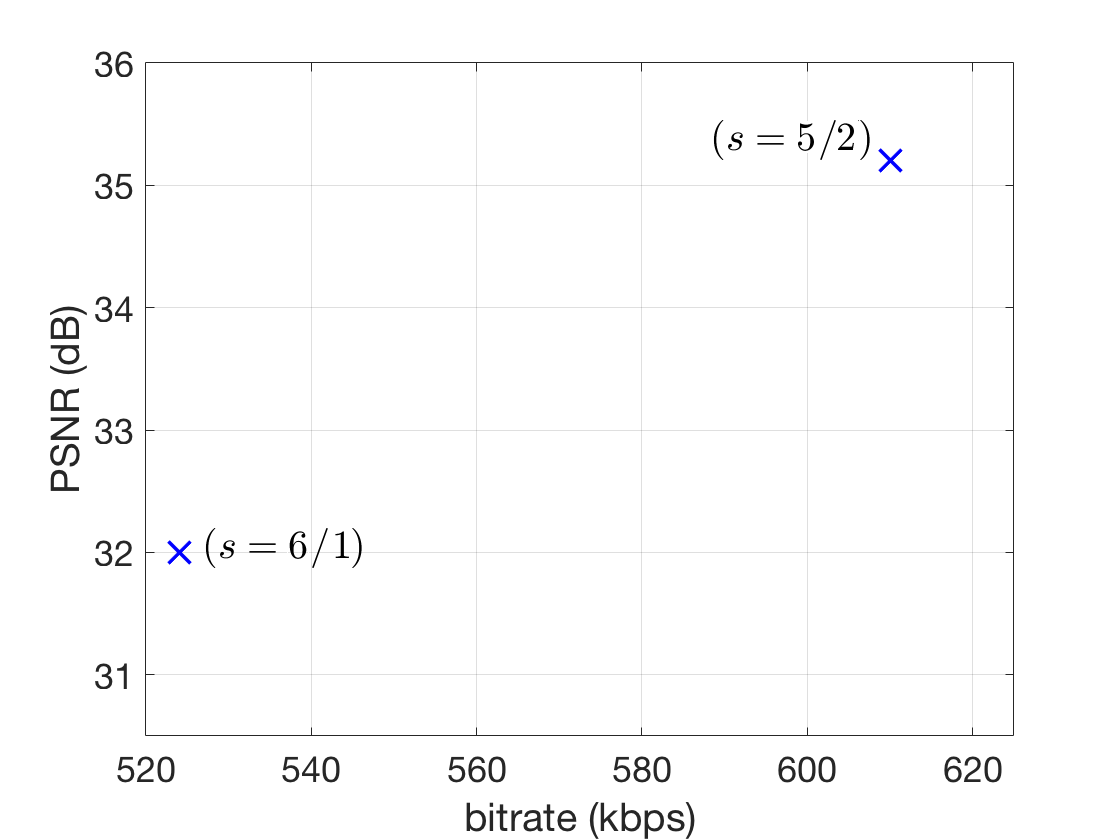}\label{fig:elimination_convexity}} ~
        \subfloat[]{\includegraphics[width= 0.24\textwidth]{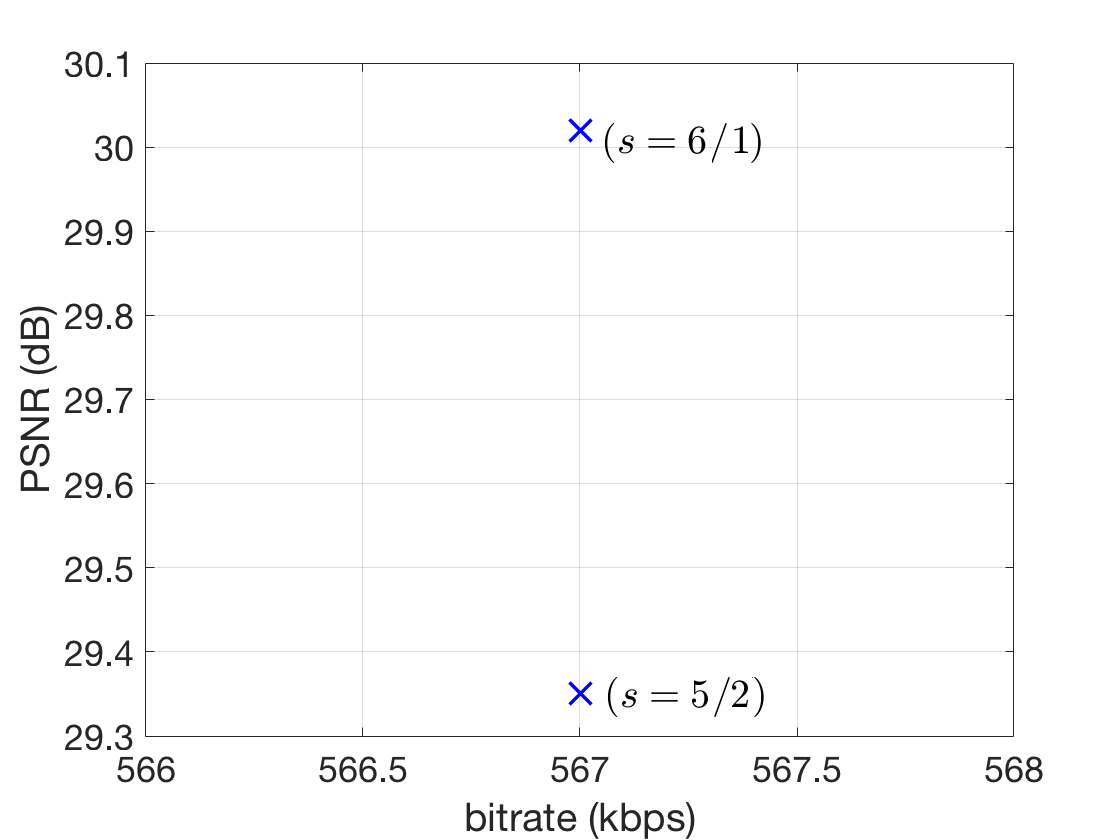}\label{fig:remapping}} ~
        \caption{Illustration of the operation of the proposed precoding mode selection algorithm during the encoding of the first 30 frames of the \textit{aspen} FHD video sequence. (a) RD points corresponding to all encodings of the first GOP with H.264/AVC for all scale factors in $\boldsymbol{S}$. (b) Remaining RD points after the pruning of the points in (a) that do not provide for a monotonically increasing RD curve. (c) RD points remaining after the elimination of the points that do not lie on the convex hull of the RD points in (b). (d) Remapping of the RD points in (c) with CBR encoding at average bitrate after the pruning process.}
        
\end{figure*}

For illustration purposes, we demonstrate in Fig. \ref{fig:footprinting}-\ref{fig:remapping} the operation of the mode selection algorithm on the first GOP (first 30 frames) of the \textit{aspen} FHD sequence when the latter is encoded with H.264/AVC with target bitrate set to 500kbps. For downscaling, we use nine scale factors $\boldsymbol{S} = \{1, 5/4, 4/3, 3/2, 2, 5/2, 3, 4, 6\}$ with  our trained multi-scale precoding network (described in Section \ref{sec:precod_networks}), while upscaling is performed using the bilinear filter. Fig. \ref{fig:footprinting} shows the set of RD points obtained after precoding the GOP for each scale factor in $\boldsymbol{S}$ and {encoding with H.264/AVC via VBV encoding with the CRF values of Table \ref{tab:crf_values}, which is representative of real-world streaming presets \cite{asan2018optimum}, and both maximum bitrate and buffer size set to 500kbps}. As shown in Fig. \ref{fig:elimination_monotonicity}, the RD points (and the corresponding precoding modes) that do not provide for a monotonically increasing rate-PSNR curve are eliminated from the RD plane. Since there are more than two points remaining after this step, we next prune out the points that do not lie on the convex hull, which leaves us with two candidate precoding modes $s=6$ and $s=5/2$. We finally re-encode the GOP, precoded with the two modes, with CBR encoding at 567kbps, which is the average bitrate of the two points in Fig. \ref{fig:elimination_convexity}.  This results in the two RD points shown in Fig. \ref{fig:remapping}. The selected precoding mode is $s^*=6$, since it renders the highest PSNR value between the two points. Notice that the CBR encoding acts as an RD remapping and leads to both modes obtaining lower PSNR values than their PSNRs under VBV  encoding. However, the absolute PSNR values are not relevant, since we are looking for the maximum PSNR under CBR; the final precoding mode is applied on the entire GOP and the encoding uses the VBV encoding mode preset. Finally, it is also of interest to notice that for this low-bitrate case, the $s=1$ case (full resolution) is immediately shown to be suboptimal, and the mode selection is left to select from the higher-quality/higher-bitrate mode of $s=5/2$ and the lower-quality/lower-bitrate mode of $s=6$, with the latter prevailing when remapping the two points into CBR encoding at their average rate.

\section{Experimental Results}
\label{sec:experiments}

In this section, we evaluate our proposed adaptive video optimization framework in scenarios with the widest practical impact. We first compare the performance of the proposed multi-scale precoding network  with the performance of standard linear downscaling filters for individual precoding modes. Then, we evaluate the entire adaptive video precoding framework by comparing it to standard video codecs via their FFmpeg open-source implementations, as well as to external highly-regarded video encoders that act as third-party anchors. {The use of FFmpeg encoding libraries instead of the reference software libraries provided by the MPEG/ITU-T or AOMedia standardization bodies allows for the use of the same VBV model architecture for all tests and corresponds to a widely-used streaming-oriented scenario found in systems deployed around the world.}

\begin{table}
\centering
\caption{CRF values for  libx264 and libx265 and speed for libvpx-vp9 under our utilized precoding modes. }
\label{tab:crf_values}
\begin{tabular}{ c  c c c}
\toprule
 & \multicolumn{3}{c}{Encoders}\\ \cmidrule(l){2-4}
 $s$ & libx264 & libx265 & libvpx-vp9 \\ 
\midrule
 1 & 23 & -  & - \\
  5/4 & 23 & - & - \\
 4/3 & 23 & 23 & speed=1 \\
 3/2 & 23 & 23 & speed=1 \\
 2 & 18 & 18 & speed=1\ \\
 5/2 & 18 & 18 & speed=1 \\
 3 & 18 & 18 & speed=1 \\
 4 & 18 & 18 & speed=1 \\
 6 & 18 & 18 & speed=1 \\
\bottomrule
\end{tabular}
\end{table}

\begin{table*}
\centering
\caption{BD-rate ($\Delta$R) and BD-PSNR ($\Delta$P) results for representative downscaling factors.}
\begin{tabular}{ c   c   c   c   c   c   c   c   c }
\toprule
\multirow{3}{*}{} & \multicolumn{4}{ c } {H.264/AVC} & \multicolumn{4}{ c }{H.265/HEVC}  \\  \cmidrule(l){2-9}
& \multicolumn{2} { c } {bicubic} & \multicolumn{2}{ c }{ Lanczos} & \multicolumn{2} { c } {bicubic} & \multicolumn{2}{ c }{ Lanczos} \\ \cmidrule(l){2-9} 
$s$ & $\Delta$R  & $\Delta$P & $\Delta$R & $\Delta$P & $\Delta$R & $\Delta$P & $\Delta$R & $\Delta$P     \\ \midrule
5/2     &   -24.70\%    &   0.61dB  &  -19.21\%  &   0.45dB  &   -25.17\%   &   0.55dB  &    -18.84\%  &  0.39dB \\
2       &   -18.85\%    &   0.56dB  &  -14.71\%  &   0.42dB  &   -19.25\%        &   0.52dB  &    -14.46\%       & 0.37dB \\
3/2     &   -17.11\%    &   0.45dB  &  -11.75\%   &   0.31dB  &  -13.18\%        &   0.32dB  &     -8.26\%   &  0.20dB \\
\bottomrule                       
\end{tabular}
\label{tab:HD_PSNR}
\end{table*}

\begin{table*}
\centering
\caption{BD-rate ($\Delta$R) and BD-VMAF ($\Delta$V) for representative downscaling factors. }
\begin{tabular}{ c   c   c   c   c   c   c   c   c }
\toprule
\multirow{3}{*}{} & \multicolumn{4}{ c } {H.264/AVC} & \multicolumn{4}{ c }{H.265/HEVC}  \\  \cmidrule(l){2-9}
& \multicolumn{2} { c } {bicubic} & \multicolumn{2}{ c }{ Lanczos} & \multicolumn{2} { c } {bicubic} & \multicolumn{2}{ c }{ Lanczos} \\ \cmidrule(l){2-9} 
$s$ & $\Delta$R  & $\Delta$V & $\Delta$R & $\Delta$V & $\Delta$R & $\Delta$V & $\Delta$R & $\Delta$V     \\ \midrule
5/2 &  -39.74\% &  7.86 &       -34.30\%        &  6.49 &  -39.73\%     &  7.03   &  -33.75\%     & 5.74 \\
2 &  -30.32\% &  5.81 & -27.57\%        &  5.18 &  -30.20\%     &  5.12 &  -27.41\%       & 4.57 \\
3/2 &     -23.21\% &  3.43 &    -21.73\%        &  3.18 &  -18.66\%     &  2.61   &  -17.67\%     & 2.46 \\
\bottomrule                       
\end{tabular}
\label{tab:HD_VMAF}
\end{table*}

\subsection{Content and Test Settings}
\label{sec:test_setup}

The test content comprises 16 FHD ($1920 \times 1080$) and 14 UHD ($3840 \times 2160$) standard video sequences in 8-bit YUV420 format from the XIPH collection\footnote{https://media.xiph.org/video/derf/  The 4K ($4096\times 2160$) sequences used were cropped to the $3840\times 2160$ (UHD)\ section of the central portion and were encoded to 8-bit YUV420 format (using x265 lossless compression) prior to encoding to produce UHD sequences. For the experiments of Section \ref{sec:eval_precod_modes}, we used only the first 240 frames of these UHD sequences as well as the standard two-pass rate control settings in FFmpeg \cite{twopassH264}. This corresponds to the typical configuration used in general-purpose video encoding. {For the experiments of Section \ref{sec:eval_adaptive_precoding}, we used the single-pass VBV encoding settings for libx264/libx265/libvpx-vp9 that correspond to OTT streaming scenarios and can be found in FFmpeg documentation \cite{VBVH264}  and the downscaling factors and CRF values of Table \ref{tab:crf_values}. }}, {which have also been used in AOMedia standardization efforts.} The FHD test content comprises the sequences  \textit{aspen}, \textit{blue\_sky}, \textit{controlled\_burn}, \textit{rush\_field\_cuts}, \textit{sunflower}, \textit{rush\_hour}, \textit{old\_town\_cross}, \textit{crowd\_run}, \textit{tractor}, \textit{touchdown}, \textit{riverbed}, \textit{red\_kayak}, \textit{west\_wind\_easy}, \textit{pedestrian\_area}, \textit{ducks\_take\_off}, \textit{park\_joy}, which have frame rates between 25fps and 50fps. The UHD sequences used in the tests are \textit{Netflix\_BarScene},  \textit{Netflix\_Boat}, \textit{Netflix\_BoxingPractice}, \textit{Netflix\_Crosswalk}, \textit{Netfix\_Dancers}, \textit{Netflix\_DinnerScene}, \textit{Netflix\_DrivingPOV}, \textit{Netflix\_FoodMarket}, \textit{Netflix\_FoodMarket2}, \textit{Netflix\_Narrator}, \textit{Netflix\_RitualDance}, \textit{Netflix\_RollerCoaster}, \textit{Netflix\_Tango}, \textit{Netflix\_TunnelFlag}, all at  60fps. 

\begin{figure*}
\centering
        \subfloat[]{\includegraphics[width=0.44\textwidth]{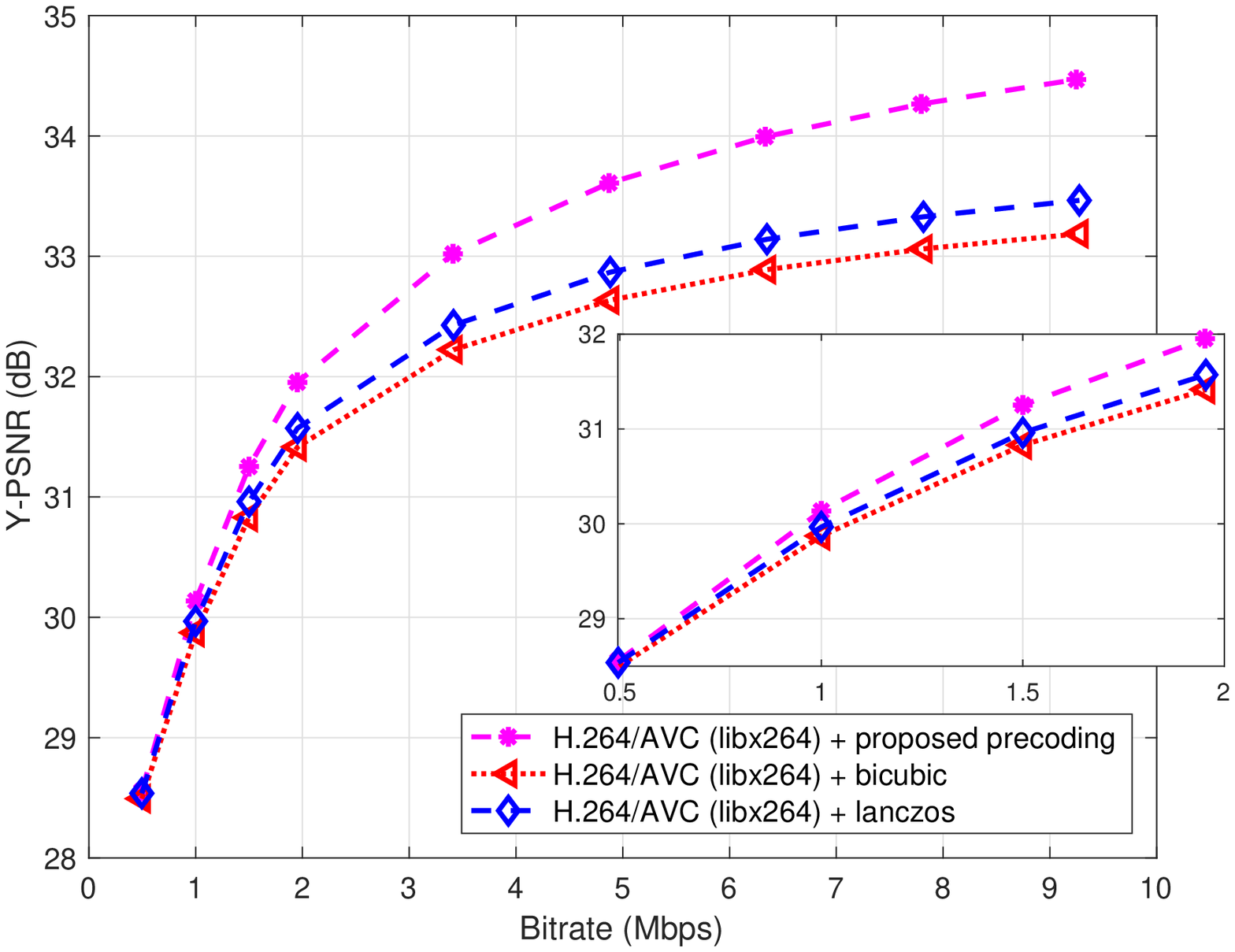}\label{fig:HD_2over5_PSNR}} ~ 
        \subfloat[]{\includegraphics[width=0.44\textwidth]{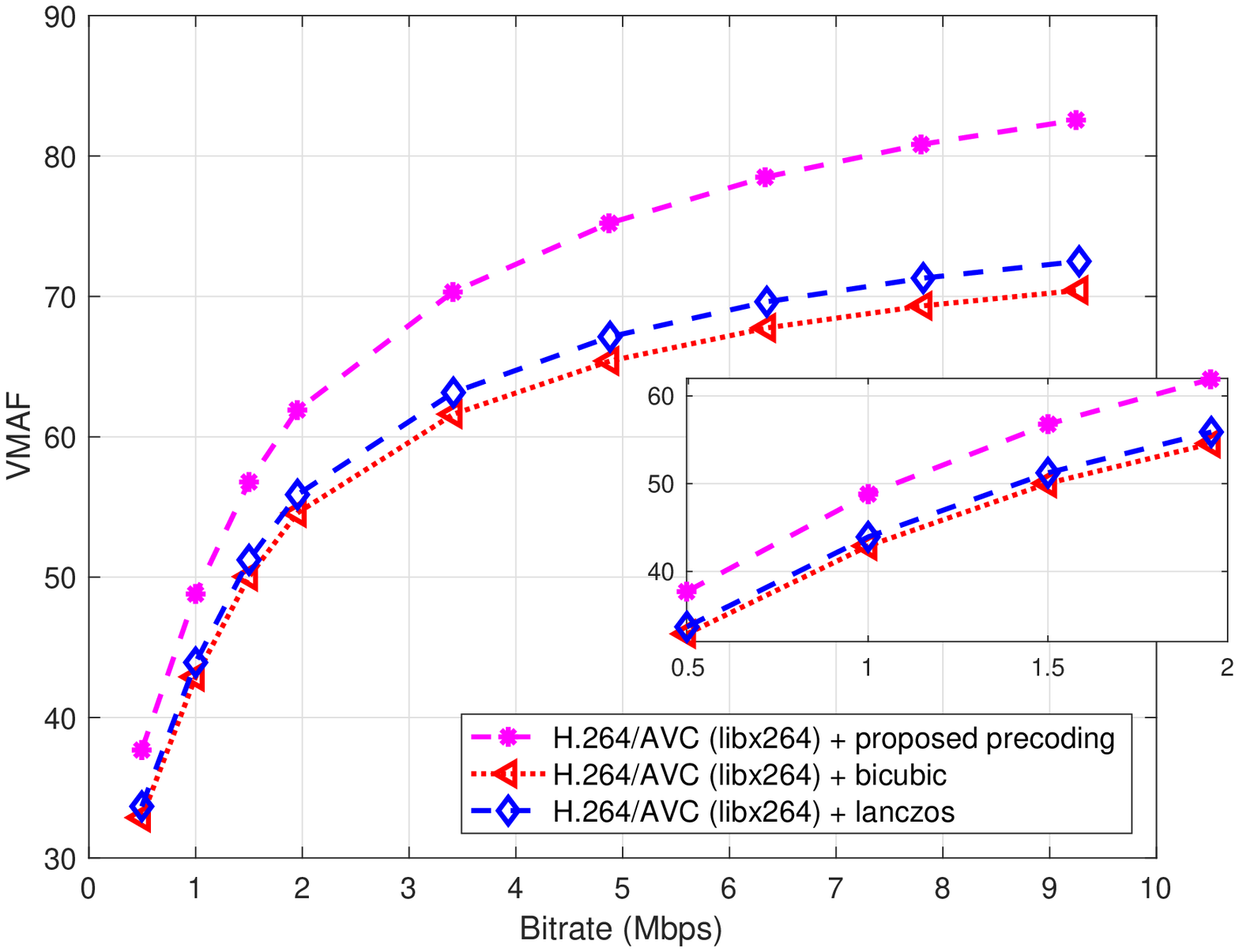}\label{fig:HD_2over5_VMAF}} 
        \caption{{Rate-distortion curves in terms of (a) PSNR and (b) VMAF for FHD content encoded with H.264/AVC and scale factor $s=5/2$.}  \label{fig:algo_illustation}}
       
\end{figure*}

Performance is measured in terms of average PSNR and average VMAF, calculated with the tools made available by Netflix \cite{li2016toward}. Average PSNR is the arithmetic mean of the PSNR values of all YUV\ channels of each frame. Similarly to PSNR, VMAF is measured per frame and the average VMAF is obtained by taking the arithmetic mean over all frames. While PSNR has been used for decades as a visual quality metric, VMAF is a relatively recent  perceptual video quality metric adopted by the video streaming community, which has been shown to correlate very well with the human perception of video quality. It is a self-interpretable $[0,100]$ visual quality scoring metric that uses a pretrained fusion approach to merge several state-of-the-art individual visual quality scores into a single metric. Both PSNR and VMAF are calculated on native resolution frames after decoding and upscaling with the bilinear filter that is supported by all video players and web browsers.

\subsection{Evaluation of Precoding Modes}
\label{sec:eval_precod_modes}

We first evaluate the performance of our proposed multi-scale precoding network against the bicubic and Lanczos filters, which are the two standard downscaling filters supported by all mainstream encoding libraries like FFmpeg. We focus  on three indicative scale factors on FHD content, opting for a very common scenario of   H.264/AVC encoding under its FFmpeg libx264  implementation. Specifically, we use the ``medium'' preset, two pass  rate control mode \cite{twopassH264}, GOP=30, and bitrate range of $0.5-10$Mbps.

BD-rate \cite{bjontegaard2001calculation} gains with respect to PSNR and VMAF are shown in Table \ref{tab:HD_PSNR} and Table \ref{tab:HD_VMAF}, respectively. Our precoding is shown to consistently outperform bicubic and Lanczos downscaling for all modes. For PSNR, its BD-rate gains ranged from 8\% to 25\%, while, for VMAF, rate reduction of 18\%-40\% is obtained. Indicative rate-distortion curves with respect to PSNR and VMAF for $s=5/2$ scaling factor are presented in Fig. \ref{fig:HD_2over5_PSNR} and Fig. \ref{fig:HD_2over5_VMAF}, showing that the proposed precoding network consistently outperforms conventional downscaling filters.  {While the gain increases at higher bitrates, substantial gain is observed in the low bitrate region as well. Specifically, for PSNR, the BD-rate and BD-PSNR gains over bicubic (Lanczos) downscaling for the 0.5-2Mbps rate region (zoomed part of the curve in Fig. \ref{fig:HD_2over5_PSNR}) are 10.54\% (6.9\%) and 0.27dB (0.18dB), respectively. For VMAF, the  BD-rate and BD-VMAF gains over bicubic (Lanczos) downscaling are 29.29\% (24.74\%) and 5.94 (4.91), respectively, for the same low bitrate region.  Example segments of a frame encoded at 5000kbps with the proposed precoding, and the Lanczos and bicubic downscaling are shown in Fig. \ref{fig:visual_illustration1} and Fig. \ref{fig:visual_illustration2}.  The improvement in visual fidelity demonstrated in the figures is also captured by the (approx.) 10-point average VMAF difference shown at the 5Mbps point of Fig. \ref{fig:HD_2over5_VMAF}. Several of the FHD video sequences encoded with a variation of the proposed precoding and H.265/HEVC\ (and the corresponding H.265/HEVC\ encoded results)  are also available for visual inspection at \href{https://www.isize.co/portfolio/demo}{www.isize.co/portfolio/demo}. They can be played with any player, since our proposal does not require any change at the streaming or client side.}

\begin{figure}
\centering
\includegraphics[width=1.0\columnwidth]{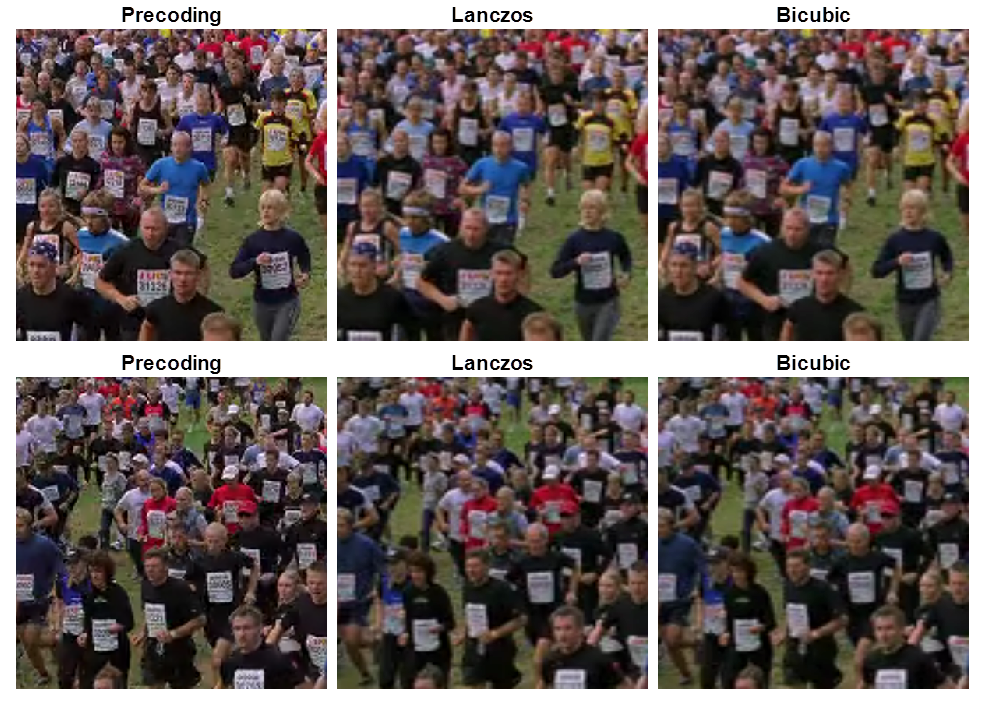}
\label{fig:precoding_2over5}
 \caption{Two segments of frame 25 of the \textit{crowd\_run}  FHD sequence encoded at 5000Kbps with the settings corresponding to Fig. \ref{fig:algo_illustation}. The precoded stream preserves the lettering and overall shapes significantly better than Lanczos and bicubic downscaling. Best viewed under magnification.\label{fig:visual_illustration1}}
\end{figure}

\begin{figure}
\centering
\includegraphics[width=1.0\columnwidth]{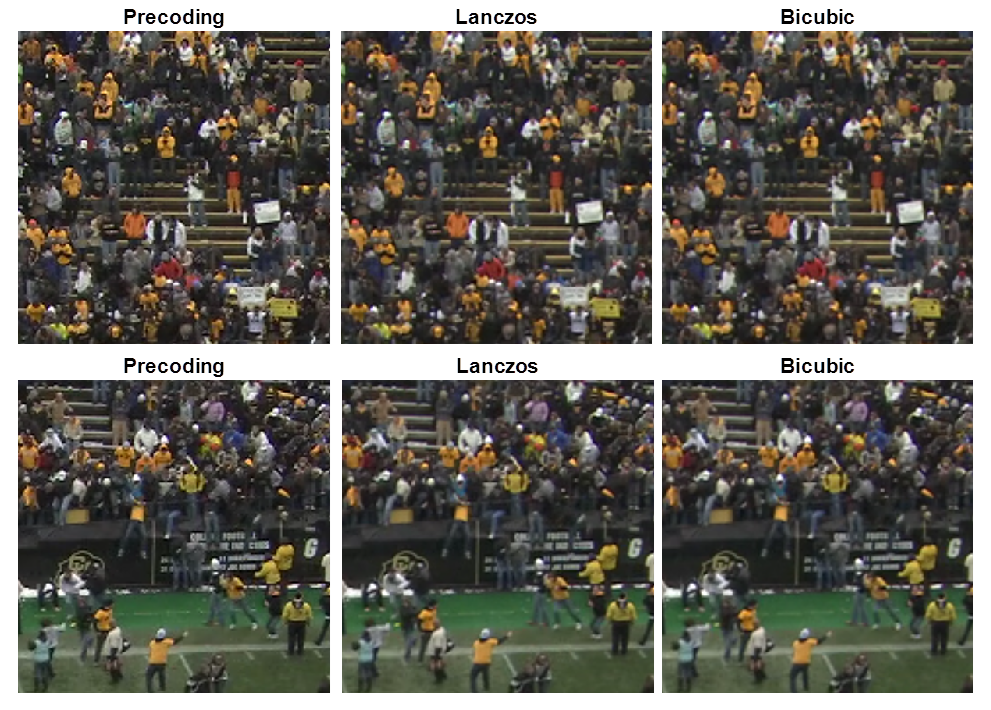}
\label{fig:bicubic_2over5}
\caption{Two segments of frame 77 of  \textit{rush\_field\_cuts}   FHD\ sequence encoded at 5000Kbps with the settings of Fig. \ref{fig:algo_illustation}. The precoded stream preserves the geometric structures significantly better than Lanczos and bicubic downscaling. Best viewed under magnification.\label{fig:visual_illustration2}}
\end{figure}

\begin{table}[t]
\centering
\caption{Average BD-rate ($\Delta$R) and BD-PSNR\ ($\Delta$P) for 16 FHD test sequences and settings described in Section \ref{sec:eval_adaptive_precoding}. The proposed precoding, iSize, in conjunction with H.264/AVC, H.265/HEVC and VP9 is evaluated against standalone H.264/AVC, H.265/HEVC and VP9 encodings as well as commercial solutions including AWS MediaConvert (AVC and HEVC) and AWS Elastic Transcoder (VP9).}
\begin{tabular}{ c   c   c   c   c   c   c  }
\toprule
\multirow{2}{*}{} & \multicolumn{2}{ c } {AVC + iSIZE} & \multicolumn{2}{ c }{HEVC + iSIZE} & \multicolumn{2}{ c }{VP9 + iSIZE}  \\  \cmidrule(l){2-7}

                                & $\Delta$R             & $\Delta$P     & $\Delta$R       & $\Delta$P     & $\Delta$R             & $\Delta$P \\
        \midrule
AVC                     & -14.80\%              & 0.72dB                & --                      & --                    & --                            & -- \\ 
HEVC            & --                            & --                    & -8.09\%         & 0.27dB                & --                            & -- \\
VP9                     & --                            & --                    & --              & --                    & -30.70\%              & 1.21dB \\
AWS             & -18.23\%              & 0.86dB                & --                    & --                      & -19.62\%              & 1.17dB \\
\bottomrule                       
\end{tabular}
\label{tab:HD_adaptive_PSNR}
\end{table}

\begin{table}[t]
\centering
\caption{Average BD-rate ($\Delta$R) and BD-VMAF ($\Delta$V) for 16 FHD test sequences and settings described in Section \ref{sec:eval_adaptive_precoding}. The proposed precoding, iSize, in conjunction with H.264/AVC, H.265/HEVC and VP9 is evaluated against standalone H.264/AVC, H.265/HEVC and VP9 encodings as well as commercial solutions including AWS MediaConvert (AVC and HEVC) and AWS Elastic Transcoder (VP9).
}
\begin{tabular}{ c   c   c   c   c   c   c  }
\toprule
\multirow{2}{*}{} & \multicolumn{2}{ c } {AVC + iSIZE} & \multicolumn{2}{ c }{HEVC + iSIZE} & \multicolumn{2}{ c }{VP9 + iSIZE}  \\  \cmidrule(l){2-7}

                                & $\Delta$R             & $\Delta$V     & $\Delta$R       & $\Delta$V     & $\Delta$R             & $\Delta$V \\ 
\midrule
AVC                     & -26.29\%              & 8.47          & --                    & --              & --                            & -- \\
HEVC                    & --                            & --                    & -15.57\%        &3.03           & --                            & -- \\
VP9                     & --                            & --                    & --                      &--                     & -25.81\%              & 6.07 \\
AWS             & -41.60\%              &12.18          & --                    &--                 & -19.52\%                      & 6.75\\
\bottomrule                       
\end{tabular}
\label{tab:HD_adaptive_VMAF}
\end{table}

\begin{table}[t]
\centering
\caption{Average BD-rate ($\Delta$R) and BD-PSNR ($\Delta$P) for 14 UHD test sequences and settings described in Section \ref{sec:eval_adaptive_precoding}. The proposed precoding, iSize, in conjunction with H.264/AVC, H.265/HEVC and VP9 is evaluated against standalone H.264/AVC, H.265/HEVC and VP9 encodings as well as commercial solutions including AWS MediaConvert (AVC and HEVC) and AWS Elastic Transcoder (VP9).}
\begin{tabular}{ c   c   c   c   c   c   c  }
\toprule
\multirow{2}{*}{} & \multicolumn{2}{ c } {AVC + iSIZE} & \multicolumn{2}{ c }{HEVC + iSIZE} & \multicolumn{2}{ c }{VP9 + iSIZE}  \\  \cmidrule(l){2-7}

                        & $\Delta$R     & $\Delta$P     & $\Delta$R     & $\Delta$P       & $\Delta$R     & $\Delta$P\\ 
 \midrule
AVC                     & -52.30\%      & 4.17dB                & --             & --                    & --                    & --    \\
HEVC            & --                    & --                    & -17.76\%      & 0.58dB          & --                    & -- \\
VP9                     & --                    & --                    & --                      & --                    & -48.82\%      & 2.03dB \\
AWS             & -47.25\%      & 3.77dB                & --                     & --                    & -36.50\%              & 2.02dB \\
\bottomrule                       
\end{tabular}
\label{tab:4K_adaptive_PSNR}
\end{table}

\begin{table}[t]
\centering
\caption{Average BD-rate ($\Delta$R) and BD-VMAF ($\Delta$V) for 14 UHD test sequences and settings described in Section \ref{sec:eval_adaptive_precoding}. The proposed precoding, iSize, in conjunction with H.264/AVC, H.265/HEVC and VP9 is evaluated against standalone H.264/AVC, H.265/HEVC and VP9 encodings as well as commercial solutions including AWS MediaConvert (AVC and HEVC) and AWS Elastic Transcoder (VP9).}
\begin{tabular}{ c   c   c   c   c   c   c  }
\toprule
\multirow{2}{*}{} & \multicolumn{2}{ c } {AVC + iSIZE} & \multicolumn{2}{ c }{HEVC + iSIZE} & \multicolumn{2}{ c }{VP9 + iSIZE}  \\  \cmidrule(l){2-7}

                                & $\Delta$R     & $\Delta$V     & $\Delta$R     & $\Delta$V       & $\Delta$R     & $\Delta$V \\ 
 \midrule
AVC                     & -46.58\%      & 15.52         & --                    & --                      & --                    & -- \\
HEVC            & --                    & --                    & -19.68\%      & 4.27                    & --                    & -- \\
VP9                     & --                    & --                    & --                      & --                    & -33.32\%      & 5.82 \\
AWS             & -68.70\%      &  35.11        &  --                   & --              & -67.77\%      & 16.65\\
\bottomrule                       
\end{tabular}
\label{tab:4K_adaptive_VMAF}
\end{table}

\subsection{Evaluation of Adaptive Precoding for Video-on-Demand Encoding }
\label{sec:eval_adaptive_precoding}

Since precoding can be applied to any codec and any video  resolution, there is a virtually unlimited range of tests that can be carried out to assess its performance on multitude scenarios of interest. Here, we focus on test conditions appropriate for highly-optimized video-on-demand (VOD) encoding systems that are widely deployed today for video delivery.  Our evaluation focuses on average bitrate-PSNR and birate-VMAF curves for our test FHD and UHD sequences and we present the results for: H.264/AVC in Fig. \ref{fig:HD_avc} and Fig. \ref{fig:4K_avc}; H.265/HEVC\ in Fig. \ref{fig:HD_hevc} and Fig. \ref{fig:4K_hevc}; and VP9 in Fig. \ref{fig:HD_vp9} and Fig. \ref{fig:4K_vp9}. {The corresponding BD-rate gains of our approach, {which we term ``iSize''}, in conjunction with each of these encoders vs. the corresponding encoder implementations, {averaged over all test sequences,} are presented in Tables \ref{tab:HD_adaptive_PSNR}-\ref{tab:4K_adaptive_VMAF}.}
For the proposed precoding method, we use footprinting with speed-up factor 5, i.e., only every 5th frame is processed during the selection of the best precoding mode, and the same encoding configuration is used as for the corresponding baseline encoder.

Regarding H.264/AVC and H.265/HEVC, we use the highly-optimized ``slower'' preset  and VBV encoding for libx264/libx265, with GOP=90, and the widely-used crf=23 configuration for VBV (see footnote 4 for further details). For our approach, we employed (per codec) the precoding modes and crf values shown in Table \ref{tab:crf_values}. To illustrate that our gains are achieved over a commercially competitive VOD encoding setup, for H.264/AVC we also include results with the high-performing AWS MediaConvert encoder\footnote{AWS tools do not support H.265/HEVC, so no corresponding benchmark is presented for that encoder from implementations external to FFmpeg. However, libx265 is well recognized as a state-of-the-art implementation and is frequently used in encoding benchmarks \cite{DeCockSPIE2016}. } using the MULTI\_PASS\_HQ H.264/AVC profile and its recently-announced high-performance QVBR mode with the default value of quality level 7.  The results of Tables \ref{tab:HD_adaptive_PSNR}-\ref{tab:4K_adaptive_VMAF} show that, against the H.264/AVC libx264 implementation, the average rate saving of our approach for both FHD and UHD resolution under both metrics (PSNR and VMAF) is {35\%}; the corresponding saving of our approach against H.264/AVC AWS MediaConvert is {44\%}.\ For H.265/HEVC libx265, the  average saving of our approach is {15\%}.     

Regarding VP9, we employed VBV  encoding with min-max rate (see more details at \cite{FFmpegVP9}), GOP=90 frames, maxrate=1.45$\times$minrate, speed=1 for lower-resolution encoding (see Table \ref{tab:crf_values}) and speed=2 for the full-resolution encoding anchor, since we only utilize downscaled versions with 6\% to 64\%\ of the video pixels of the original resolution. Additional bitrate reduction may be achievable by utilizing two-pass encoding in libvpx-vp9, but we opted not to use VBV encoding to make our comparison balanced with the VP9 implementation provided by the AWS Elastic Transcoder, which was used as our external benchmark for VP9. The settings of the Elastic Transcoder jobs were based on the built-in presets\footnote{https://docs.aws.amazon.com/elastictranscoder/latest/developerguide/preset-settings.html}, which we customized to match the desired output video codec, resolution, bitrate, and GOP size, and we set the framerate according to the input video framerate. Such customization is necessary because the built-in presets do not follow the input video parameters and they serve mainly as boilerplates.  The results of Tables \ref{tab:HD_adaptive_PSNR}-\ref{tab:4K_adaptive_VMAF} show that, against the VP9 libvpx-vp9 implementation, the average rate saving of our approach for both FHD and UHD resolution under both metrics (PSNR and VMAF) is {35\%}; the corresponding saving of our approach against VP9 AWS Elastic Transcoder is {36\%}. 

\begin{figure*}
\centering
\subfloat[]{\includegraphics[width=0.44\textwidth]{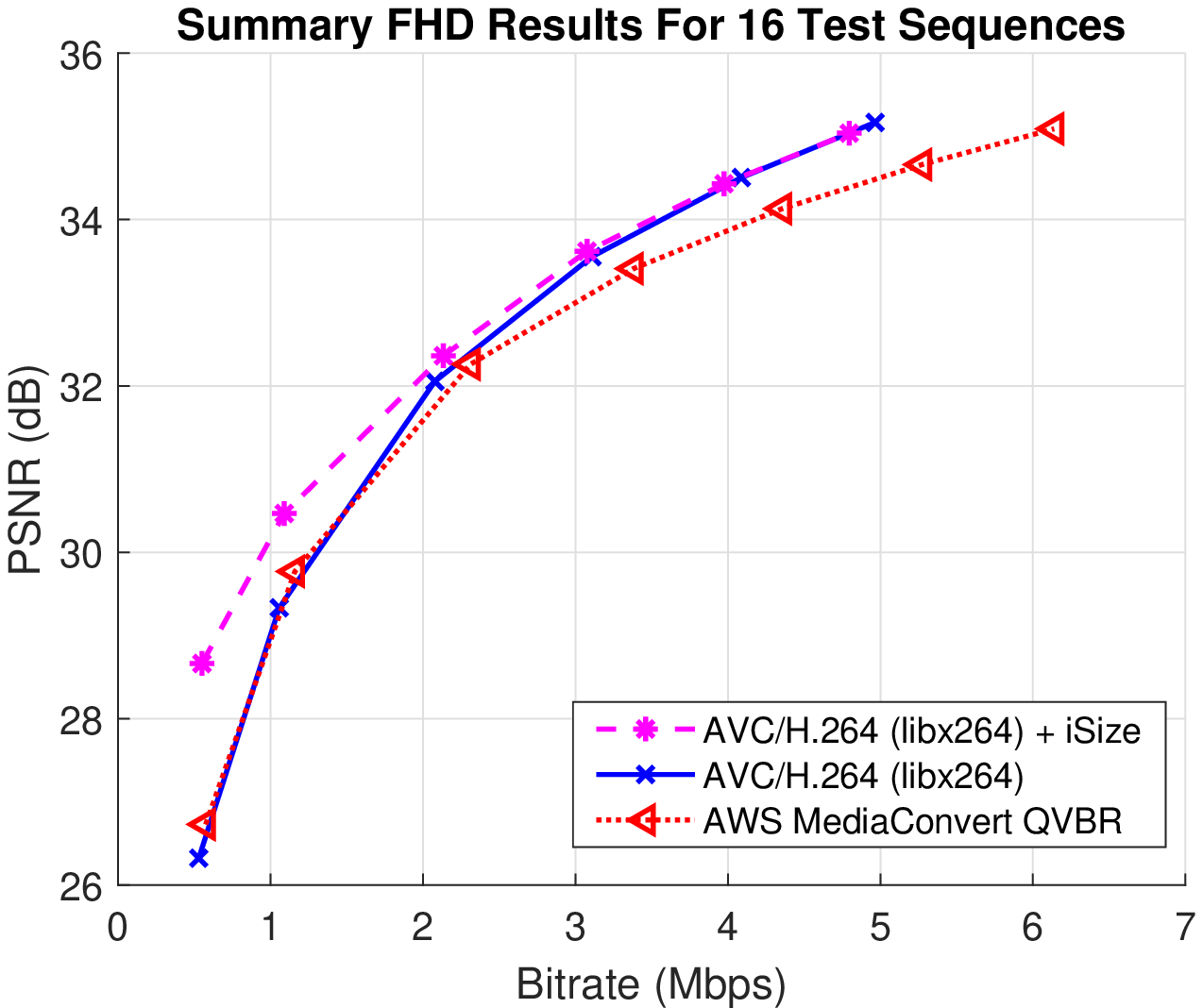}} ~
\subfloat[]{\includegraphics[width=0.44\textwidth]{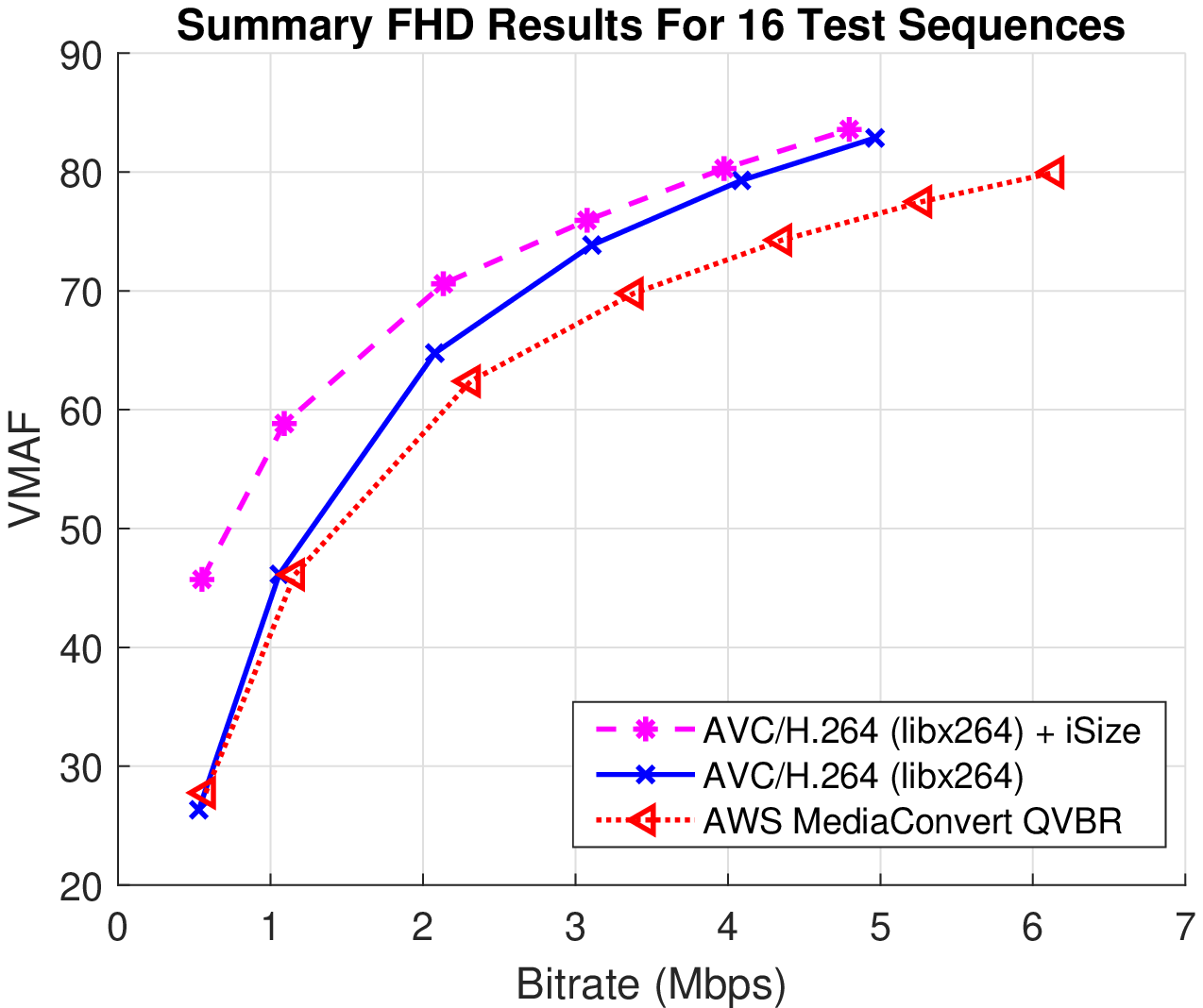}}
\caption{Performance comparison of H.264/AVC encoding with proposed adaptive precoding versus standalone H.264/AVC encoding and AWS MediaConvert H.264/AVC encoder (QVBR\ mode)\ on FHD content: (a) PSNR and (b) VMAF. }
\label{fig:HD_avc}
\end{figure*}

\begin{figure*}
\centering
\subfloat[]{\includegraphics[width=0.44\textwidth]{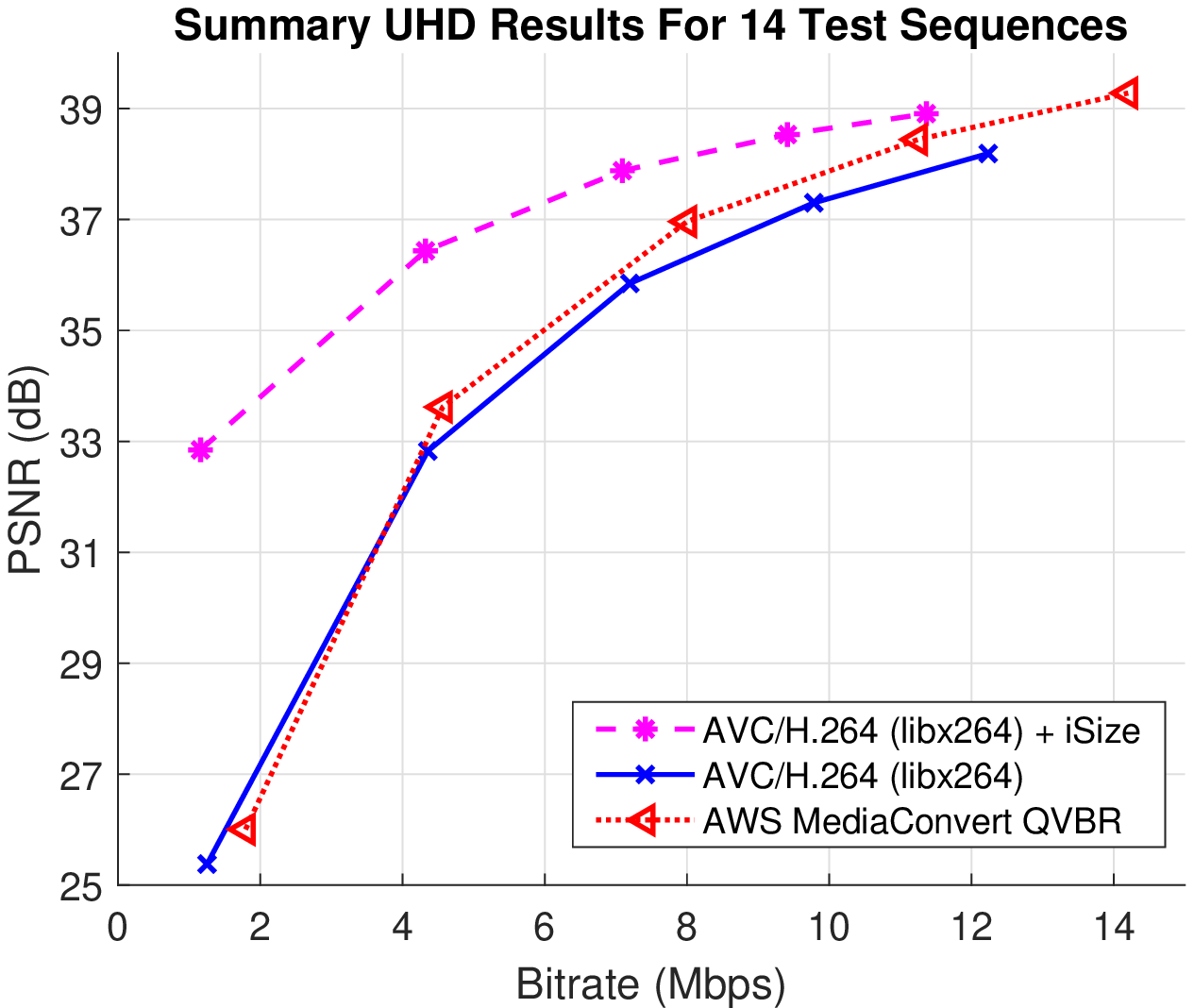}} ~
\subfloat[]{\includegraphics[width=0.44\textwidth]{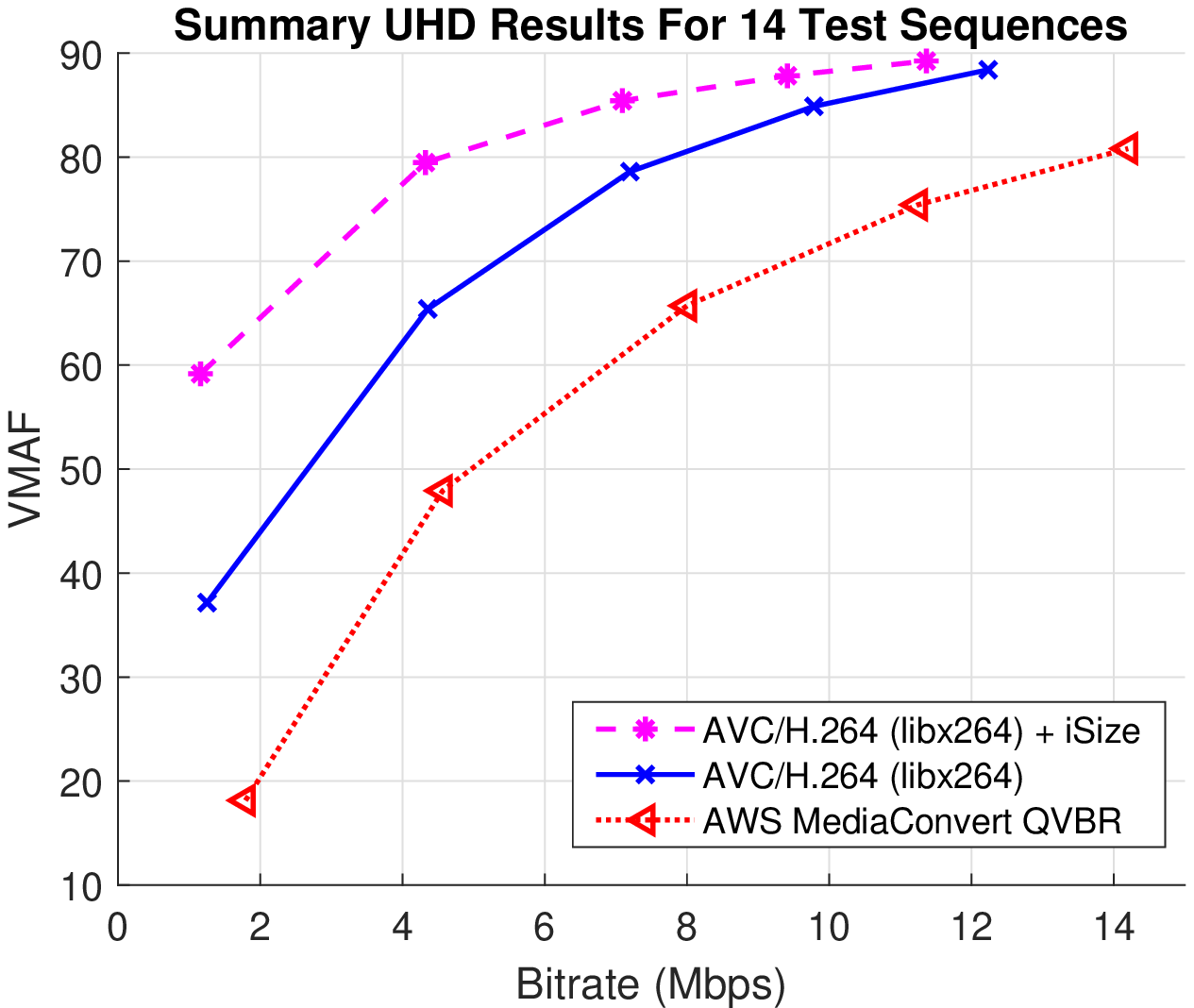}}
\caption{Performance comparison of H.264/AVC encoding with proposed adaptive precoding versus standalone H.264/AVC encoding and AWS MediaConvert H.264/AVC encoder (QVBR\ mode)\ on UHD content: (a) PSNR and (b) VMAF. }
\label{fig:4K_avc}
\end{figure*}

\begin{figure*}
\centering
\subfloat[]{\includegraphics[width=0.44\textwidth]{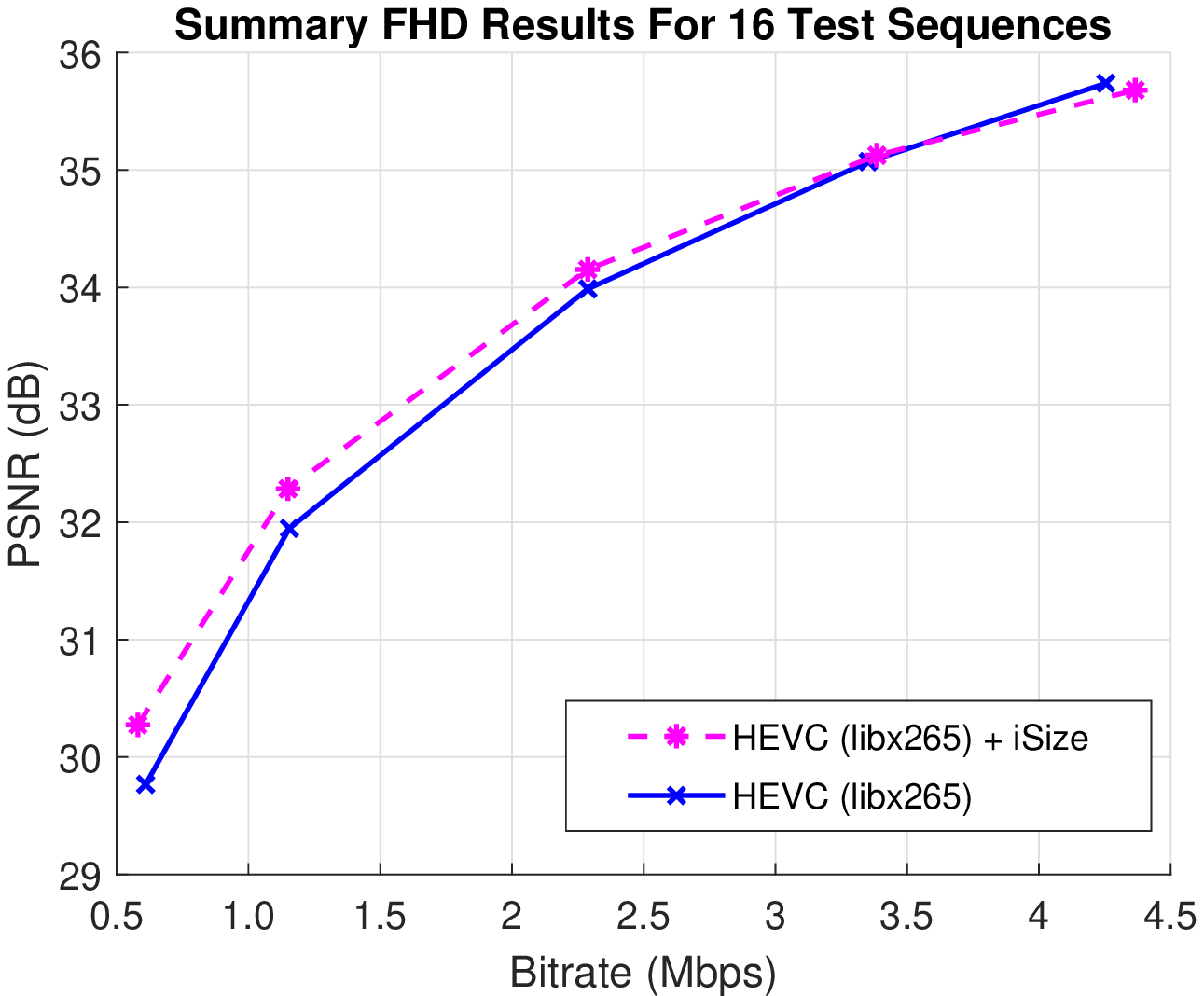}} ~
\subfloat[]{\includegraphics[width=0.44\textwidth]{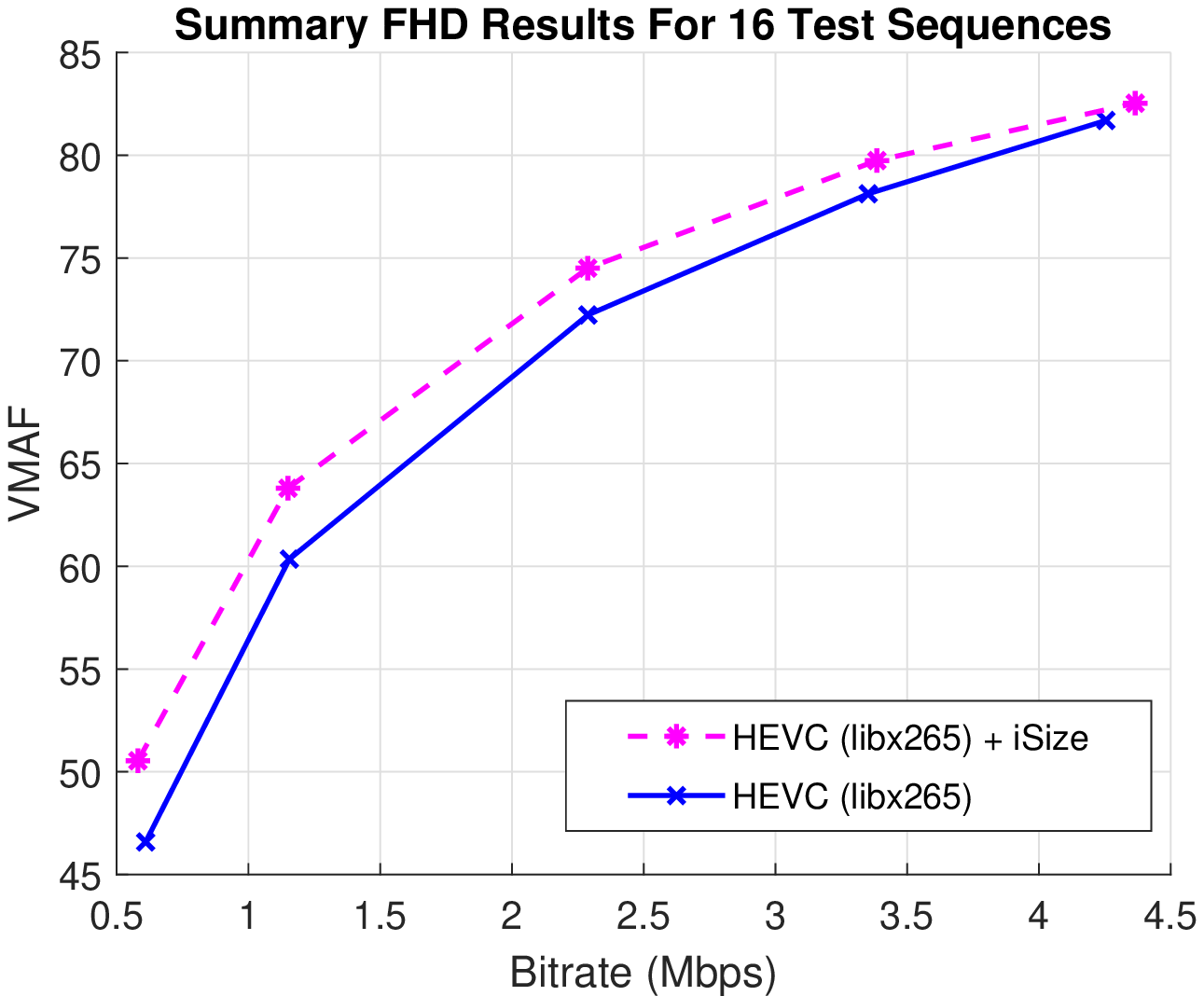}}
\caption{Performance comparison of H.265/HEVC encoding with proposed adaptive precoding versus standalone H.265/HEVC encoding on FHD content: (a) PSNR and (b) VMAF. }
\label{fig:HD_hevc}
\end{figure*}

\begin{figure*}
\centering
\subfloat[]{\includegraphics[width=0.44\textwidth]{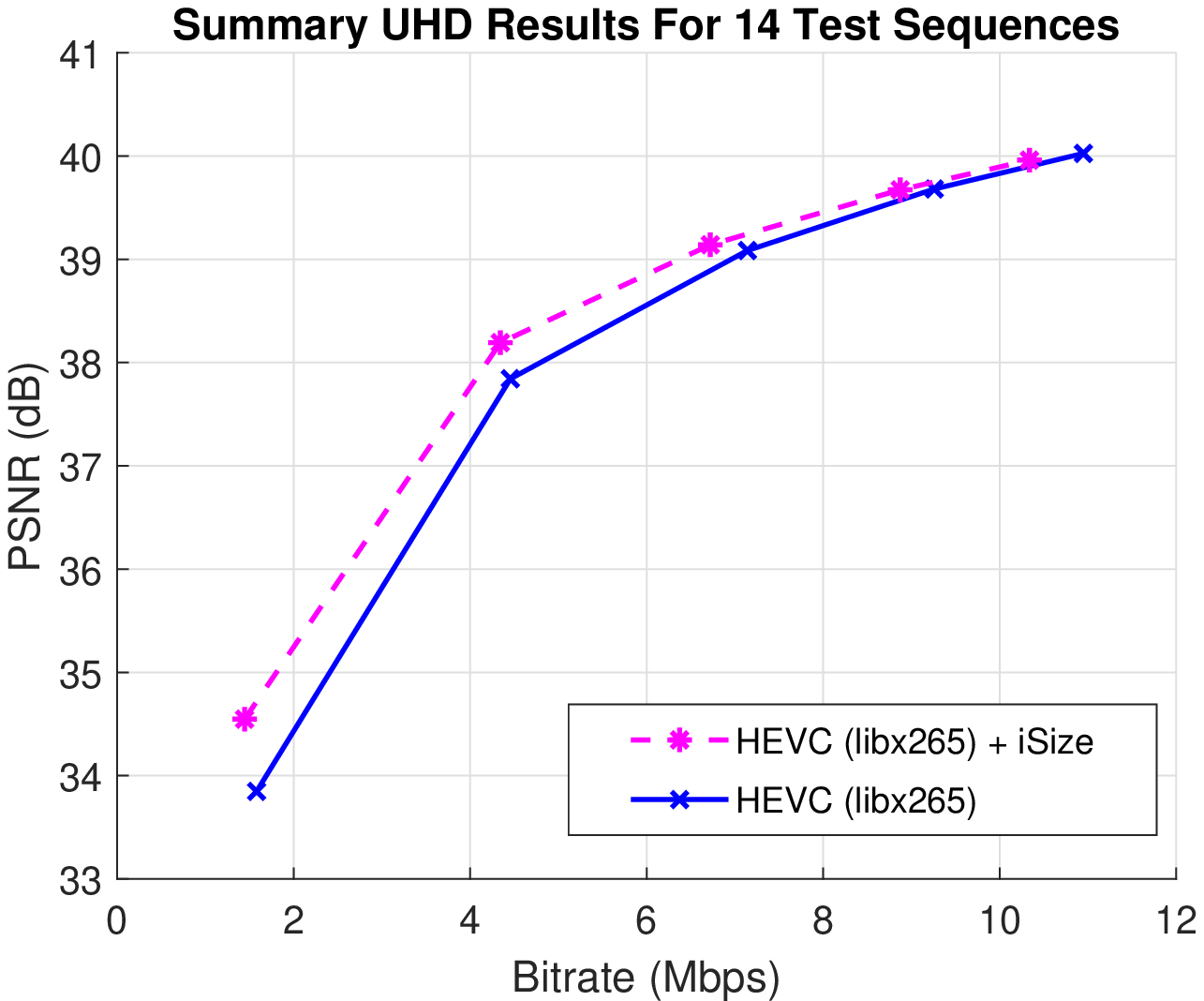}} ~
\subfloat[]{\includegraphics[width=0.44\textwidth]{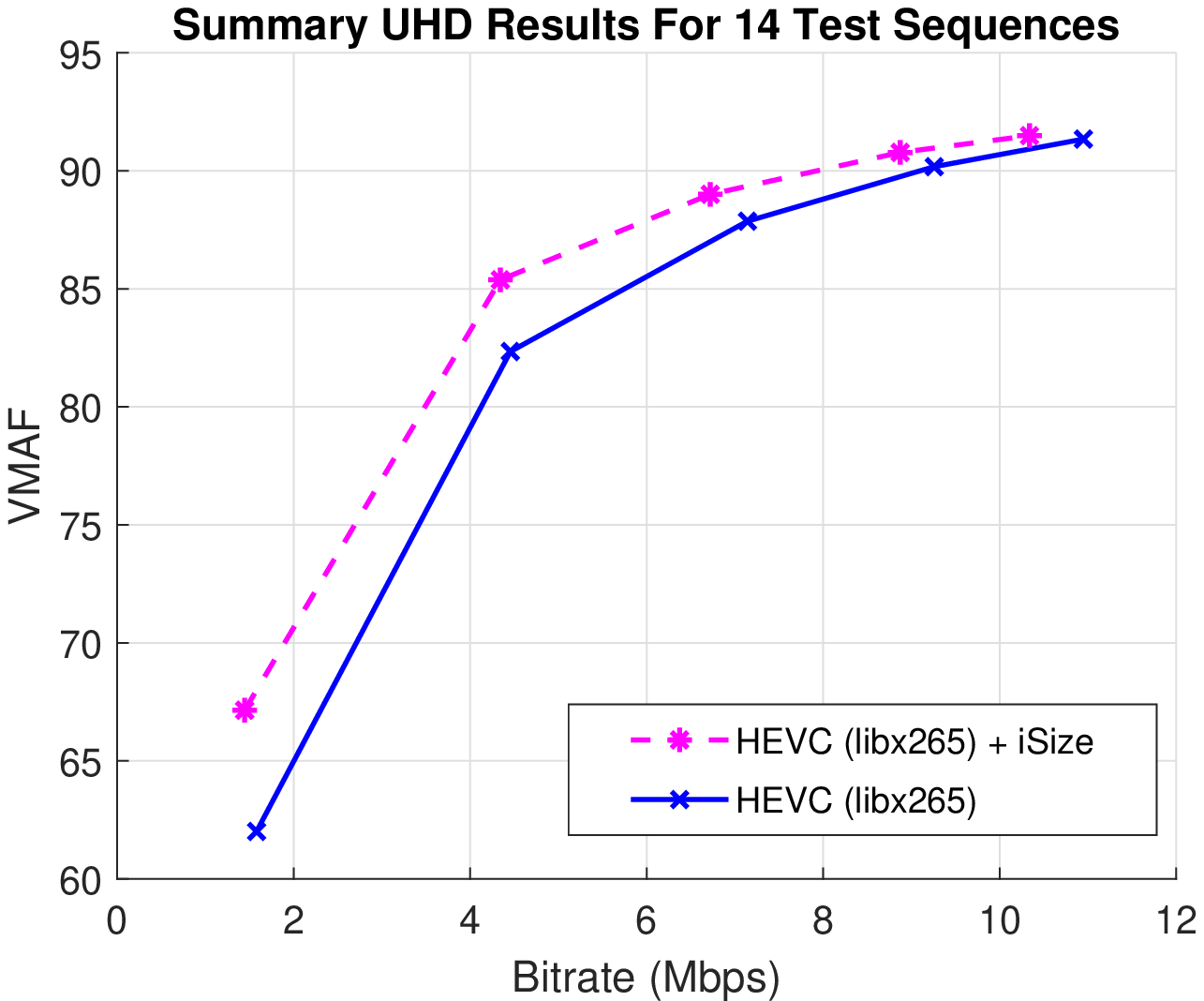}}
\caption{Performance comparison of H.265/HEVC encoding with proposed adaptive precoding versus standalone H.265/HEVC encoding  on UHD content: (a) PSNR and (b) VMAF. }
\label{fig:4K_hevc}
\end{figure*}

\begin{figure*}
\centering
\subfloat[]{\includegraphics[width=0.44\textwidth]{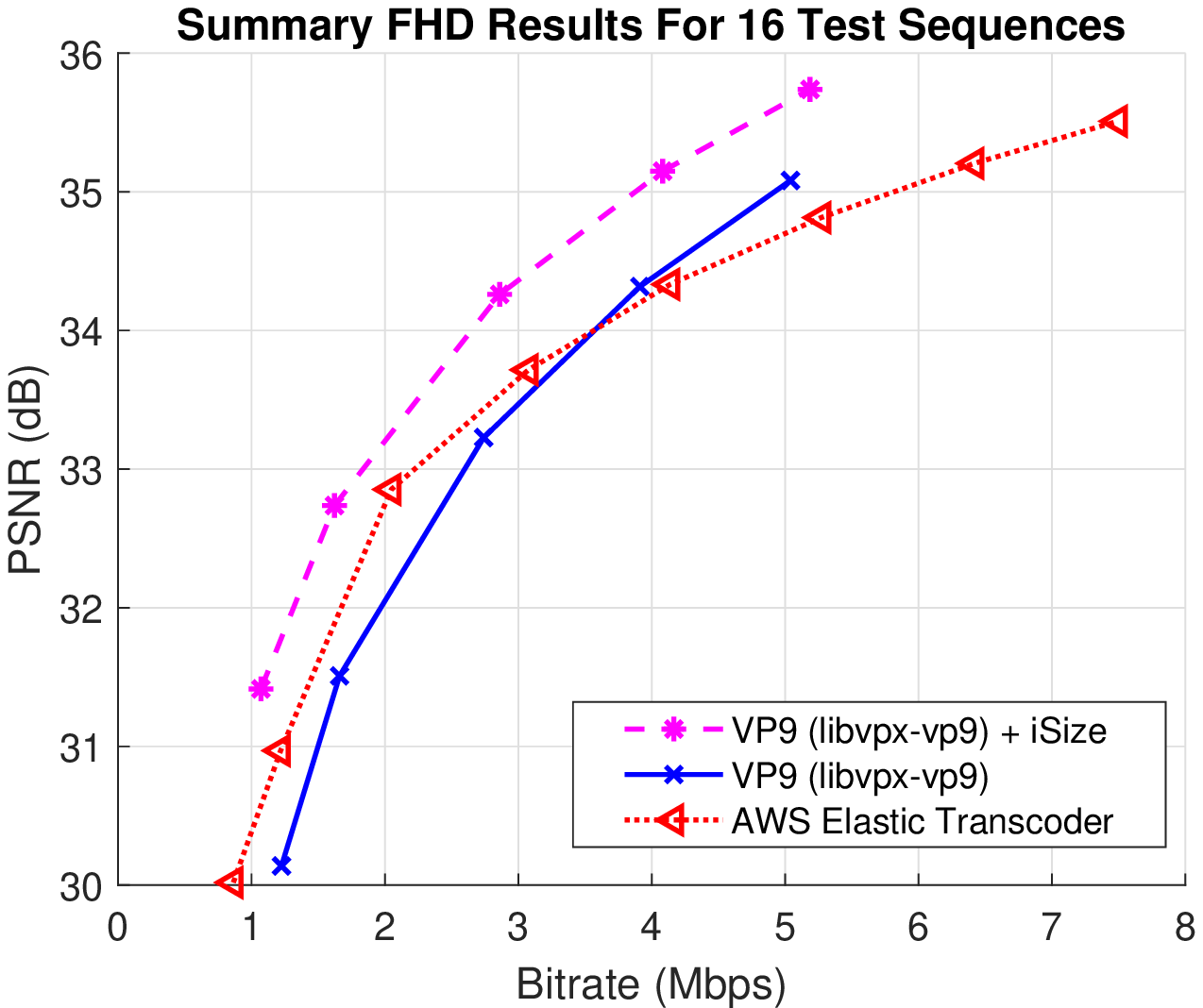}} ~
\subfloat[]{\includegraphics[width=0.44\textwidth]{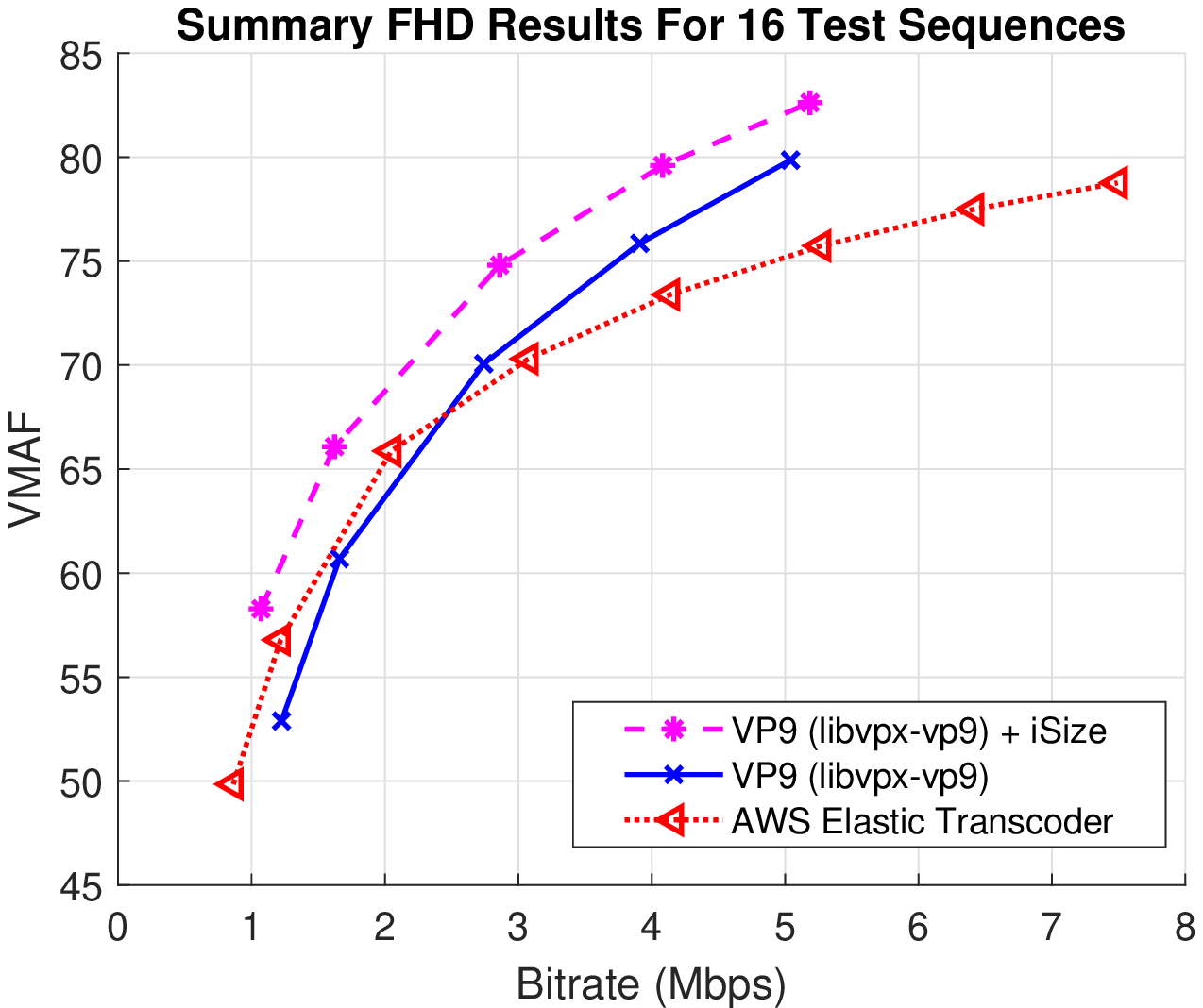}}
\caption{Performance comparison of VP9 encoding with proposed adaptive precoding versus standalone VP9 encoding and AWS Elastic Transcoder VP9 encoder on FHD content: (a) PSNR and (b) VMAF }
\label{fig:HD_vp9}
\end{figure*}

\begin{figure*}
\centering
\subfloat[]{\includegraphics[width=0.44\textwidth]{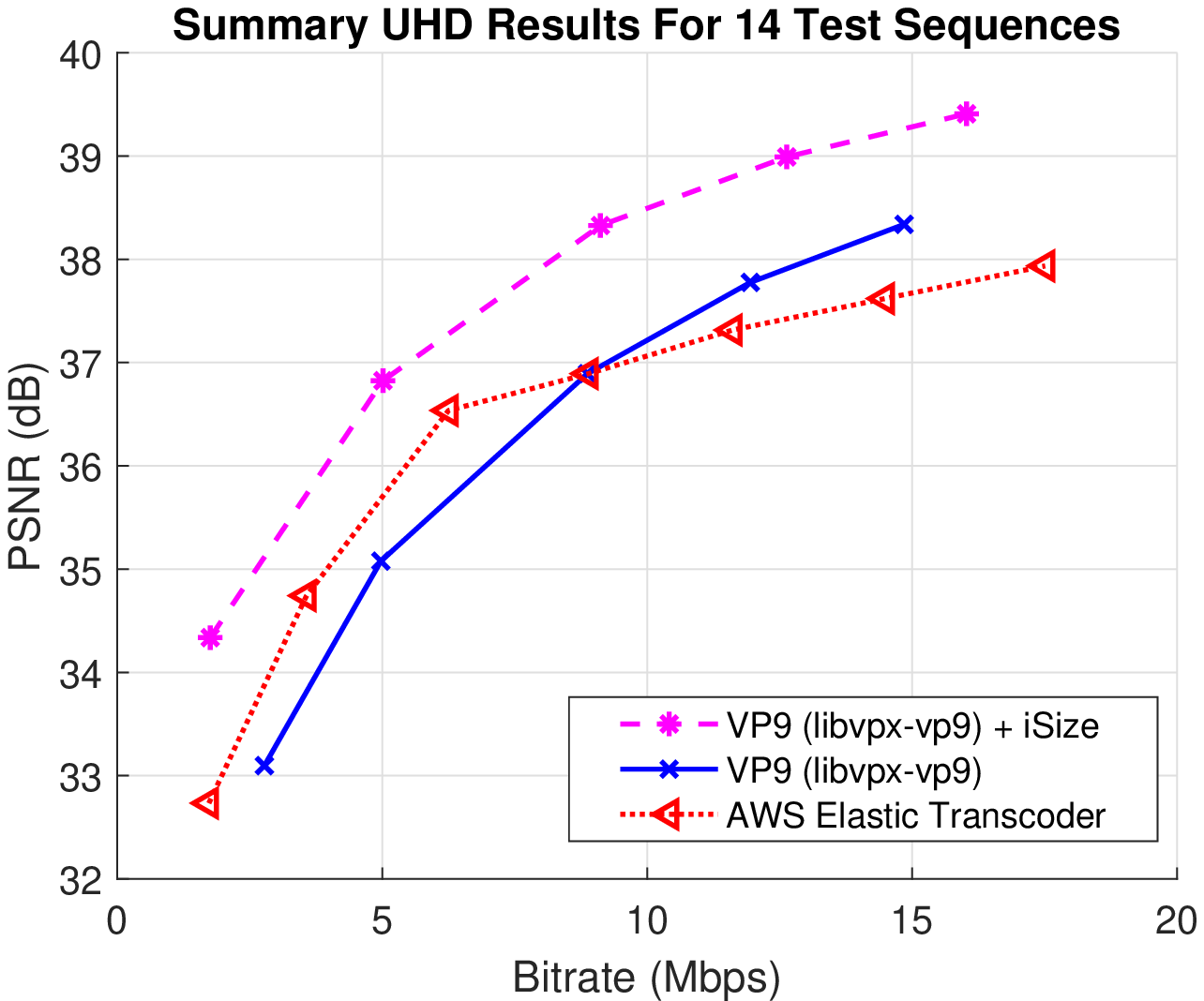}} ~
\subfloat[]{\includegraphics[width=0.44\textwidth]{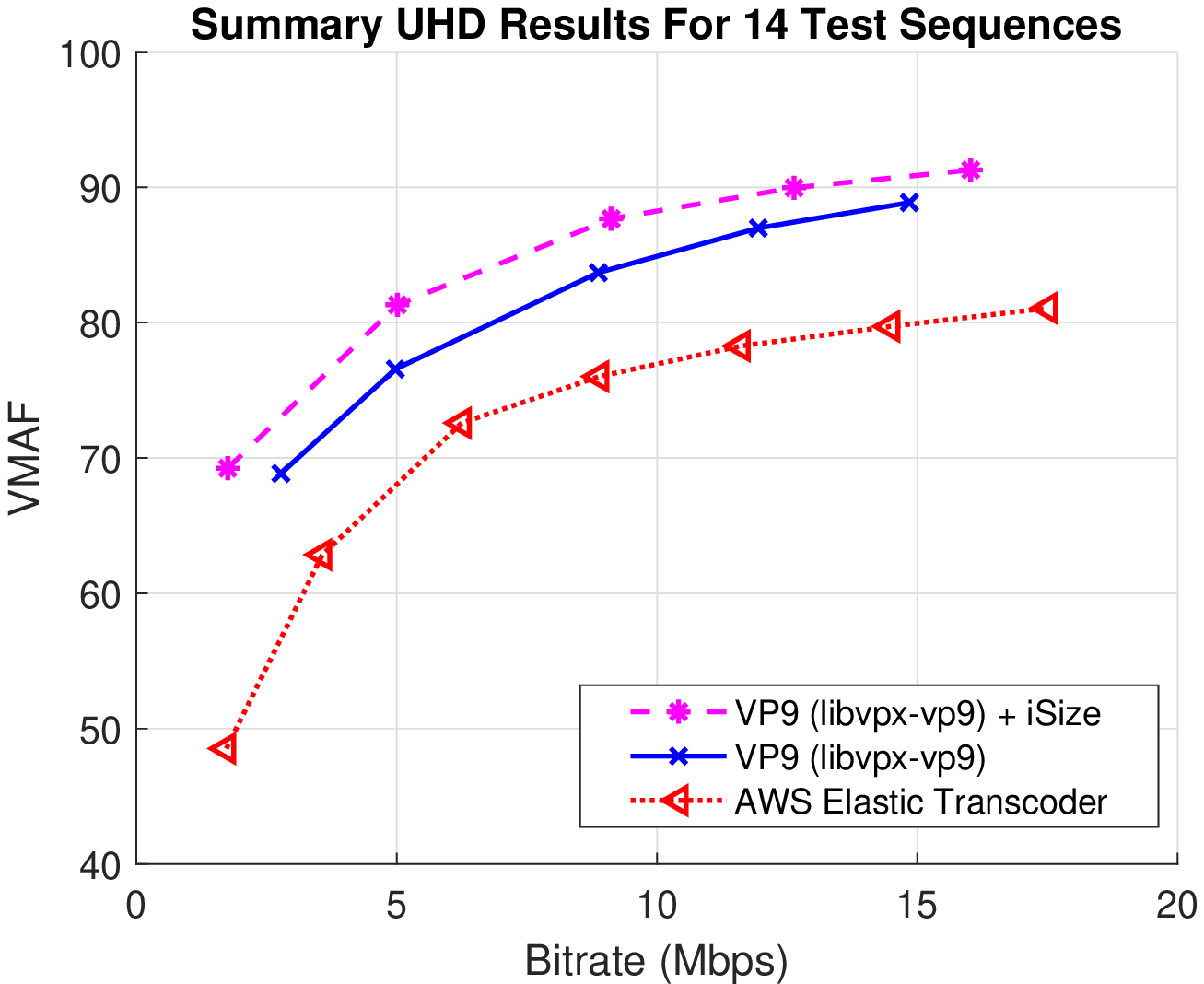}}
\caption{Performance comparison of VP9 encoding with proposed adaptive precoding versus standalone VP9 encoding and AWS Elastic Transcoder VP9 encoder on UHD content: (a) PSNR and (b) VMAF. }
\label{fig:4K_vp9}
\end{figure*}

{
\subsection{Further Comparisons and Discussion}
\label{sec:further_comp_disc}

\subsubsection {Evaluation of adaptive precoding on HD (720p) content} To examine the performance of our approach for lower-resolution inputs, we carried out a smaller-scale evaluation of our proposed adaptive precoding method on four $1280\times720$ HD video sequences, namely \textit{ducks\_take\_off}, \textit{in\_to\_tree}, \textit{old\_town\_cross} and \textit{park\_joy} from the XIPH collection. The average BD-rate gains of our approach versus the standalone H.264/AVC encoding for bitrates in the region of 0.5-2.5Mbps and same settings as in Section \ref{sec:eval_adaptive_precoding} were found to be 3.01\% and 0.11dB for PSNR, and 2.27\% and 0.68 for VMAF. These gains are modest in comparison to those obtained for FHD and UHD content. This indicates that our proposal is more suitable for high-resolution content.  

\subsubsection{Evaluation of adaptive precoding in conjunction with the current VVC Test Model (VTM)} Beyond comparisons with existing standards, to show that our proposed deep video precoding can offer gains even over upcoming video coding standards, we evaluate our method in conjunction with the current VTM (version 6.2rc1). As the utilized VTM is very slow (2-10min per frame on single-core CPU execution), for this evaluation, we limited our tests to seven FHD sequences from the XIPH collection (\textit{aspen, controlled\_burn, old\_town\_cross, crowd\_run, rush\_field\_cuts, touchdown, tractor}) and two bitrates: 1.8mbps and 3mbps. The utilized settings were as provided in the VTM software{\footnote{https://vcgit.hhi.fraunhofer.de/jvet/VVCSoftware\_VTM}, with the default rate control enabled and set to the target bitrates and IntraPeriod is set to 64 frames}. The results are summarized in Table \ref{tab:vvc}. We can see that rate reduction of 8\%-9\% is offered at slight increase of PSNR and VMAF. This first evaluation illustrates that our proposal can achieve gains even against the current VTM model of the Joint Video Experts Team. We would also like to note that the VVC standard is still under development and it is expected that these results will change depending on the features that will be included in the final version of the standard.  

\subsubsection{Evaluation against other CNN-based downscaling frameworks}  To compare the core CNN designs of our precoding framework against recent work on CNN-based downscaling, we investigate the performance of our proposed precoding network under fixed downscaling ratio of $s=2$ followed by bilinear upscaling on four standard test image datasets: Set5, Set14, BSDS100 and Urban100. Our benchmarks comprise bicubic and CNN-based \cite{LiTIP2019} downscaling coupled with bicubic and Lanczos upscaling filters that are more complex than our bilinear upscaling. As summarized in Table \ref{tab:image_comparison},  with the exception of Li \textit{et al}. \cite{LiTIP2019}  on BSDS100 followed by Lanczos upscaling, our approach outperforms all reference methods on all datasets, and achieves this result with the very lightweight bilinear upscaling at the client side. In addition, our framework has significantly lower complexity than Li \textit{et al.} \cite{LiTIP2019}, as  for an $1920\times 1080$ input frame and $s=2$, our precoding network requires only 3.38G MACs and 5.5K parameters over all scales compared to 153G MACs and 30.6K parameters required by Li \textit{et al.} \cite{LiTIP2019}.

\subsubsection{Impact of edge preservation loss} The purpose of the edge preservation loss in Eq. \eqref{eq:loss_func_m} is to ensure structural preservation for the non-integer downscaling ratios that utilize bilinear downscaling instead of a stride. We explore the impact of the edge preservation loss by computing average PSNR and SSIM over the DIV2K validation set for  $\lambda \in \{0,0.5,2,5\}$ and indicative scaling factors. The results are reported in Table \ref{tab:ablation}. Without the edge preservation loss $(\lambda = 0)$, the average PSNR and SSIM for scaling factor $s=3/2$ that utilizes bilinear downscaling are 35.86dB and 0.957, respectively. Increasing lambda to 0.5, the PSNR and SSIM increase to 36.6dB and 0.962, respectively. A similar increase is exhibited for scaling factor $s=4/3$, where PSNR increases from 37.96dB to 38.23dB.  However, we notice that as the weight increases, the metrics saturate and there can be a detrimental effect on integer ratios, such as $s=4$, where the mean absolute error (MAE) and feed-through from higher scaling factors is sufficient in ensuring fidelity to the input frame. Therefore, in order to ensure a balance in performance over all scaling factors, we set $\lambda=0.5$ in all our results. 
}

\begin{table}[t]
\centering
\caption{{Average PSNR, VMAF and obtained bitrate for VVC encoding (VTM v.6.2rc1) with proposed precoding versus standalone VVC encoding on FHD content.}}
\begin{tabular}{ c   c   c   c   c    }
\toprule
\multirow{2}{*}{} & \multicolumn{2}{ c } {VVC+iSize} & \multicolumn{2}{ c }{VVC}  \\ 
\midrule
PSNR            &27.22dB                &28.18dB                &27.15dB         &28.04dB        \\
VMAF            &94.06          &96.57                  &91.48                  &95.72  \\
Bitrate (Mbps)          &1.636                  &2.748          &1.796                   &2.988  \\
\bottomrule                       
\end{tabular}
\label{tab:vvc}
\end{table}

\begin{table*}[t]
\centering
\caption{{Evaluation of the proposed precoding neural network coupled with bilinear upscaling filter against bicubic and CNN-based downscaling methods paired with different upscaling filters in terms of reconstruction quality (PSNR) on standard image datasets for scaling factor $s=2$.   }}
\begin{tabular}{c c c c c c c c}
\toprule 
\multirow{2}{*}{} & Bicubic $\downarrow$ &$\mathrm{CNN-CR}^{Sep}$ \cite{LiTIP2019} $ \downarrow$ & Bicubic $\downarrow$ &$\mathrm{CNN-CR}^{Sep} $ \cite{LiTIP2019} $ \downarrow$ & Bicubic $\downarrow$ &$\mathrm{CNN-CR}^{Sep} $ \cite{LiTIP2019} $ \downarrow$  & Our precoding  $\downarrow$ \\
 & Bilinear $\uparrow$ & Bilinear $\uparrow$ & Bicubic $\uparrow$ & Bicubic $\uparrow$ & Lanczos $\uparrow$ & Lanczos $\uparrow$& Bilinear $\uparrow$  \\ \midrule 
Set5            & 32.26         & 33.55         & 33.67         & 33.36         & 33.72   & 33.37         &\textbf{34.81}         \\ 
Set14   & 28.97         & 30.27 & 30.01         & 30.30 & 30.05 & 30.32         &\textbf{30.81}  \\  
BSDS100 & 28.70         & 29.80         & 29.57         & 29.86 & 29.61         & \textbf{29.87}  &29.67          \\  
Urban100 & 25.60        & 26.99         & 26.56         & 27.09 & 26.60         & 27.11   &\textbf{27.65}         \\ \midrule \midrule
Average         & 27.38         &28.63  &28.32  &28.70  &28.36  &28.72  &\textbf{28.94} \\      
\bottomrule
\end{tabular}
\label{tab:image_comparison}
\end{table*}

\begin{table*}[t]
\centering
\caption{{Effect of the edge preservation loss on PSNR/SSIM for indicative scaling factors on the DIV2K validation dataset.}}
\begin{tabular}{ c   c   c   c   c   c   c  c  c  }
\toprule
\multirow{2}{*}{} & \multicolumn{2}{ c } {$\lambda=0$} & \multicolumn{2}{ c }{$\lambda=0.5$} & \multicolumn{2}{ c }{$\lambda=2$}  & \multicolumn{2}{ c }{$\lambda=5$} \\
\cmidrule(l){2-9}
$s$     &PSNR           &SSIM           &PSNR           &SSIM           &PSNR           &SSIM           &PSNR           &SSIM \\
\midrule        
4/3     &37.96dB                &0.97247                &38.23dB                &0.97384                &38.21dB                &0.97367                &38.13dB                &0.97331\\
3/2     &35.86dB                &0.95689                &36.61dB                &0.96223                &36.61dB                &0.96216                &36.60dB                &0.96211\\
3       &29.88dB                &0.84640                &29.82dB                &0.84634                &29.77dB                &0.84569                &29.74dB                &0.84548\\
4       &28.47dB                &0.79431                &28.38          &0.79346                &28.28dB                &0.79204                &28.19dB                &0.79088\\
\bottomrule                       
\end{tabular}
\label{tab:ablation}
\end{table*}

\begin{table*}[t]
\centering
\caption{Average runtime (msec per GOP on CPU) for representative downscaling factors and multiple encoding bitrates. }
\begin{tabular}{c c c c c c c c c}
\toprule 
\multirow{2}{*}{} & \multicolumn{2}{c}{Precoding} & \multicolumn{2}{c}{AVC} & \multicolumn{2}{c}{HEVC} & \multicolumn{2}{c}{VP9} \\\cmidrule(l){2-9}
$s$ & FHD & UHD & FHD & UHD & FHD & UHD & FHD & UHD \\ \midrule 
1 & 0 & 0 & 1451 & 4928 & 11002 & 54016 & 3072 & 594400\\ 
3/2 & 671 & 4067 & 745 & 1982 & 7634 & 32565 & 2107 & 318445\\  
2 & 775 & 4468 & 529 & 1854 & 7476 & 31786 & 1763 & 221522\\  
5/2 & 703 & 4408 & 358 & 1008 & 6292 & 14692 & 1015 & 156259\\
\bottomrule
\end{tabular}
\label{tab:runtime}
\end{table*}

\subsection{Runtime Analysis for Cloud-based VOD Encoding}
\label{sec:runtime}

Since VOD encoding configurations are typically deployed over a cloud implementation, it is of interest to benchmark the complexity of encoding with our deep video precoding modes versus the corresponding plain encoder.  Table \ref{tab:runtime} presents benchmarks for typical precoding scales on an AWS t2.2xlarge instance with each precoding and encoding running in two threads on two of the Intel Xeon 3.3GHz CPUs. The results correspond to the average processing of each GOP (comprising 90 frames and excluding I/O). By comparing the standard encoding time per video coding standard (scale 1) with our precoding and encoding time for the remaining scales, we see that, as scale increases, the encoding time reduces by up to a factor of five, while precoding time remains quasi-constant. The  rate savings versus downscaling by these factors using bicubic and Lanczos filters are shown in Table \ref{tab:HD_PSNR} and Table \ref{tab:HD_VMAF} (for FHD content). These results indicate that, especially for the case of complex encoding standards like H.265/HEVC\ and VP9 that require long encoding times for high-quality VOD streaming systems, the combination of precoding with the appropriate ratio may allow for a more efficient realization on a cloud platform with substantial reduction in rate (or improvement in quality)  than using linear filters. In this context, precoding effectively acts as a data-driven pre-encoding compaction mechanism in the pixel domain, which allows for accelerated encoding, with the client linearly upscaling to the full resolution and producing high quality video.

\section{Conclusion}
\label{sec:conclusions}
We propose the concept of deep video precoding based on  convolutional neural networks, with the current focus being on downscaling-based compaction under DASH/HLS adaptation. A key aspect of our approach is that it remains backward compatible to existing systems and does not require any change for the streaming, decoder, and display components of a VOD\ solution. Given that our approach does not alter the encoding process, it offers an additional optimization dimension going beyond content-adaptive encoding and codec parameter optimization. Indeed, experiments show that it brings benefits on top of such well-known optimizations: under high-performing two-pass and VBV-based FHD and UHD video encoding, {our precoding offers 8\%-52\%  bitrate saving versus leading AVC/H.264, HEVC and VP9 encoders, with lower gains offered for HD (720p) content. An early-stage evaluation against the VVC Test Model v6.2rc1 showed that our approach may also be beneficial for advanced encoding frameworks currently under consideration by video coding standardization bodies. In addition, a comparison against a state-of-the-art CNN-based downscaling framework and bicubic or Lanczos upscaling showed that our proposal offers a better downscaler even when the less-complex bilinear filter is used for upscaling at the client side.} The compaction features of our solution ensure that, not only bitrate is saved, but also that video encoding complexity reduction can be achieved, especially for HEVC and VP9 VOD\ encoding. {Future work will consider how to extend the notion of precoding beyond adaptive streaming systems by learning to adaptively preprocess video inputs such that they are optimally recovered by current decoders under specified perceptual quality metrics}.
\bibliographystyle{IEEEtran}

\end{document}